\documentclass[12pt]{iopart}

\usepackage{iopams}
\usepackage{graphicx}
\usepackage{amssymb}
\usepackage{xcolor}
\usepackage{ulem}
\newcommand\la{\langle}
\newcommand\ra{\rangle}

\begin{document}
\title[Quantum resetting in continuous measurement induced dynamics of a qubit]{Quantum resetting in continuous measurement induced dynamics of a qubit}

\author{Varun Dubey$^1$, Raphael Chetrite$^2$, Abhishek Dhar$^1$}

\address{$^1$ 
International Centre for Theoretical Sciences, Tata Institute of Fundamental Research, Bengaluru 560089, India,\\ $^2$ Universit\'e C\^ote d’Azur, CNRS, LJAD, Parc Valrose, 06108
NICE Cedex 02, France}

\eads{varun.dubey@icts.res.in, raphael.chetrite@unice.fr and abhishek.dhar@icts.res.in}

\begin{abstract}
We study the  evolution of  a two-state system that is monitored continuously but with interactions with the detector tuned so as to avoid the Zeno affect. The system is allowed to interact with a sequence of prepared probes. The post-interaction probe states are measured and this leads to a stochastic evolution of the  system's state vector, which can be described by a single angle variable. The system's effective evolution consists of a deterministic drift and a stochastic resetting to a fixed state at a rate that depends on the instantaneous state vector. The detector readout is a counting process.  We obtain analytic results for the distribution of number of detector events and the time-evolution of the probability distribution.  Earlier work on this model found  transitions in the form of the steady state  on increasing the measurement rate. Here we study transitions seen in the dynamics. As a spin-off we obtain, for a general stochastic resetting process with diffusion, drift and position dependent jump rates, an exact and general solution for the evolution of the probability distribution.
\end{abstract}

\section{Introduction\label{sec:Intro}}
The problem of repeated measurements on  quantum systems is of great interest in the context of monitoring and controlling its time evolution and in the context of answering questions such as that of the time of arrival.  Of particular interest is the situation where a system is coupled to a probe and  repeated measurements are performed on the probe. An obviously interesting question is as to what these measurements on the probe can tell us about the system. A number of recent experiments have looked at the trajectories of quantum systems subjected to repeated measurements~\cite{guerlin2007,murch2013,roch2014,minev2019}. General discussions of measurements and quantum trajectory theory can be found in Refs.~\cite{carmichael1993,Gisin,Barchielli1982,Jacobs2006,wiseman_milburn_2009,breuer2002theory,raimond2006,Patel_Kumar}.

A quantum system that is continuously monitored via direct measurements remains frozen in its state. This is the well known quantum Zeno effect~\cite{misra1977,degasperis1974,teuscher2004,Asher}. The Zeno freezing can be avoided under the scheme of  indirect measurements  where the system is allowed to interact for a period $\tau$ with a probe, with an interaction strength that scales as $\tau^{-1/2}$, and then  the state of the probe is measured projectively~\cite{bauer2012}.  Depending upon the result of this measurement, one obtains partial information about the state of the object system. If this process is repeated with identically prepared probes, then the Zeno efect is avoided and  system evolves stochastically. It can be shown~\cite{carmichael1993,Gisin,Barchielli1982,Jacobs2006,wiseman_milburn_2009,breuer2002theory,raimond2006,Brun2002}  that, for a two-state object system interacting with a sequence of identically prepared probes (which are also two-state systems), the state of the object system evolves via a stochastic  Schr\"odinger equation with jumps. Furthermore, when the interaction strength between the object system and the probes scales as $\tau^{-1/2}$, the reduced density matrix of the  system evolves via a Lindblad equation.
Any given stochastic trajectory of the wavefunction corresponds to the system's evolution for a particular outcome of the measurement sequence. Averaging over these outcomes corresponds to the case of blind measurements and the entire information of the system's evolution is contained in the reduced density matrix.

The basis of the current work is the model described in Ref.~\cite{Snizhko}. In this work, the authors have considered a measurement problem on a two-state system similar to the one described in the last paragraph. The principal conclusion is that upon variation of the relative strength  $\lambda$ (defined below, see ~Eq.(\ref{eq:NoClickNonHerm})) of measurement, the system exhibits transitions which mark various stages in the onset of the quantum Zeno effect. Note that usual Zeno effect refers to the phenomena whereby a system's dynamics gets frozen as a result of  continuous measurements on it. This Zeno effect is avoided with the choice of interaction strength scaling as $\tau^{-1/2}$.  But what Ref.~\cite{Snizhko} finds for their model is that  in the limit of infinite measurement strength there is again a freezing of the dynamics. Interestingly, signatures of this freezing appear even at large but finite mesaurement strengths, with parts of the Hilbert space becoming inaccessible --- this is referred to as Zeno effect appearing in stages.

Following ~\cite{Snizhko}, we model the detector readouts as a counting process and investigate the onset of the Zeno regime in the counting statistics of the readout process. We note that similar investigations have been carried out in ~\cite{Li2014}. However, in the model we consider, the counting process has a stochastic intensity~\cite{Hawkes1971} given by the rate function $\alpha_t$ in Eq.~(\ref{eq:RateF}). Our calculations reveal that for $\lambda=2$, the mean count $\mathbb{E}[N_{t}]$ of the counting process exhibits a topological transition as remarked in ~\cite{Snizhko}. Measurement induced entanglement transitions and topological phase transitions based on Zeno physics have gained considerable attention, in particular we note the recent studies~\cite{li2018,TurkeshiPhysRevB.103.224210,biella2021,nahum2021,ippoliti2021}. These transitions have been identified with presence of  exceptional points ~\cite{GopalKrishnan} in the spectrum of non-Hermitian Hamiltonian which evolves the quantum state under continuous measurement and post-selection.  In ~\cite{HugoPhysRevE.98.022129}, the spectral approach is employed to investigate the properties of Markov processes that are reset to a fixed state at times picked from an exponential distribution. ~\cite{MingantiPhysRevA.100.062131,MingantiPhysRevA.101.062112} consider exceptional points of non-Hermitian Hamiltonians as well Liouvillians governing open system dynamics.

We point out that the stochastic dynamics of our qubit  can be interpreted as  the overdamped motion of a particle in a tilted periodic potential with a  resetting of the position to a particular point, at a  rate that depends on the particle  position.  Stochastic resetting  has been widely studied in the classical context~\cite{evans2011a,evans2011b,pal2016,evans2020} but there are few studies in the quantum context~\cite{mukherjee2018,yin2022,TurkeshiPhysRevB.105.L241114}. Our study provides a simple example where stochastic resetting in a quantum system appears  naturally as a result of measurements. As our second main result, we use the renewal approach to compute the exact time-dynamics of the probability distribution of the wavefunction.

The plan of this paper is as follows. In Sec.~\ref{sec:Basic}, we describe the basic setup and the measurement protocol, discuss the emergence of the stochastic Schr\"odinger equation and summarize known results from earlier work. We also discuss the Bloch sphere representation of the qubit and the particular simplification that occurs for the system we study. In  Sec.~\ref{sec:CountStat} we present the  calculation to obtain explicit expressions for the generating function for the number of clicks and from it the mean number of clicks.  Sec.~\ref{sec:SteadyState} contains the calculation for the time-evolution of the system using a renewal approach.  We then discuss a second approach based on the non perturbative formula for  the resolvent (Green's function).  
In Sec.~\ref{sec:spectral} we present some results on spectral properties of the probability evolution operator and use it to write another solution for the time evolution.    
We conclude in Sec.~\ref{sec:Conc}.

\section{Basic setup and summary of earlier work}
\label{sec:Basic}

\subsection{Definition of the model and dynamics}
Consider a $2-$level system $\mathcal{S}$ whose Hilbert space $\mathcal{H_S}$ is spanned by the vectors
\begin{equation}
\label{eq:linspan}
|\psi_{0} \ra=\left[{\begin{array}{c}
	1\\
	0\\
	\end{array}}\right],\qquad |\psi_{1}\ra=\left[{\begin{array}{c}
	0 \\
	1 \\
\end{array}}\right].
\end{equation}
The system evolves with the Hamiltonian
\begin{equation}
H_{\mathcal{S}}=\left[{\begin{array}{cc}
0 & \gamma_{0}\\ \gamma_{0} & 0
\end{array}}\right]=\gamma_{0}\sigma_{x}
\end{equation}
where $\gamma_{0}$ is a positive  frequency.  $\sigma_{x}, \sigma_{y}, \sigma_{z}$ represent the Pauli matrices.  At any instance $t$, the state of $\mathcal{S}$ is given by the normalized vector
\begin{equation}
|\psi(t)\ra=a(t)|\psi_{0}\ra+b(t)|\psi_{1}\ra=\left[{\begin{array}{c}
a(t)\\b(t)
\end{array}}\right].
\end{equation}
At this instance, $\mathcal{S}$ is allowed to interact with another $2-$level system $\mathcal{D}$ for a short time interval $\tau$. The Hilbert space $\mathcal{H_D}$ is spanned by $\{\chi_{0},\chi_{1}\}$ defined similarly as in Eq.~(\ref{eq:linspan}). At the start of the interaction, $\mathcal{D}$ is assumed to be in the state $\chi_{0}$. The combined state of the system $\mathcal{S}$ and the detector $\mathcal{D}$ is the uncorrelated vector

\begin{equation}
|\Psi(t)\ra=|\psi(t)\ra\otimes |\chi_{0}\ra
\end{equation}
in the tensor product space $\mathcal{H}=\mathcal{H_S}\otimes\mathcal{H_D}$. We adopt the convention that for states or operators in $\mathcal{H}$, the first factor corresponds to $\mathcal{S}$ and the second factor to $\mathcal{D}$ in all summands. The state $\Psi(t)$ evolves in the interval $\tau$ by the Hamiltonian
\begin{equation}
\label{eq:CombinedEvol}
H=H_{\mathcal{S}}\otimes I + \sqrt{\frac{\gamma}{\tau}}\pi_{1}\otimes\sigma_{y},
\end{equation}
where we note that the interaction part of the Hamiltonian is scaled as ${1}/{\sqrt{\tau}}$ and  $\gamma$ is a non-negative  coupling frequency.
In Eq.~(\ref{eq:CombinedEvol}), the projector $\pi_{1}=|\psi_{1}\ra \la \psi_{1}|$ and $I$ is the identity operator. It follows that the combined state  after the interval $\tau$ is given by
\begin{eqnarray}
\eqalign{|\Psi(t+\tau)\ra=\exp\left[-\rmi \tau H\right] |\Psi(t)\ra \nonumber\\
\fl =|\psi(t) \ra \otimes |\chi_{0}\ra+ (-\rmi\tau)\Big[\left(H_{S}-\rmi\frac{\gamma}{2}\pi_{1}\right) |\psi(t)\ra\Big]\otimes |\chi_{0}\ra 
-\rmi\sqrt{\gamma\tau}[\pi_{1} |\psi(t)]\ra \otimes[\sigma_{y} |\chi_{0}\ra]+\mathcal{O}(\tau^{\frac{3}{2}}).} 
\end{eqnarray}
Now a projective measurement in the basis $|\chi_{0}\ra$,$|\chi_{1}\ra$  \footnote{Note that the fact that the associated projectors $\pi_{0},\pi_{1}$ are rank $1$, which is natural in this bidimensional $\mathcal{H_D}$, is in fact for a multidimensional $\mathcal{H_D}$  the main hypothesis which permits to conserve the purity of the system state and to write,  as in the following,  an equation for the pure state instead of the (impure) density matrix.} is performed to determine the state of the detector. If the detector is found to be in the state $|\chi_{0}\ra$, then the un-normalized state $|\tilde{\psi}(t+\tau)\ra$ of the system, up to first order in $\tau$ is given by
\begin{equation}
\label{eq:Noclick}
\eqalign{|\widetilde{\psi}(t+\tau)\ra &= \left[I-\rmi\tau\left(H_{s}-\rmi\frac{\gamma}{2}\pi_{1}\right)\right]|\psi(t)\ra },
\label{Monday}
\end{equation}
The probability of this event, i.e, of the readout to be $\chi_{0}$, up to first order in $\tau$ is  given by
\begin{equation}
p_{0} = 1-\gamma\tau \la\psi|\pi_1|\psi\ra=1-\gamma\tau|b(t)|^2.
\label{Tuesday}
\end{equation}
If the readout is $\chi_{1}$, then the un-normalized state and the  probability of the readout are
\begin{eqnarray}
\label{eq:collapse}
|\widetilde{\psi}(t+\tau)\ra =\sqrt{\gamma \tau} \pi_1 |\psi(t)\ra  \\
\qquad p_{1} =\gamma\tau \la\psi|\pi_1|\psi\ra= \gamma\tau|b(t)|^2.
\end{eqnarray}
This completes description of one measurement cycle. Subsequently the object system is coupled to another detector initialized in $\chi_{0}$ and the process is repeated sequentially. Every time a detector is measured to be in the state $\chi_{1}$ corresponds to a 'click'.

In the limit $\tau = dt \to 0$ the stochastic evolution of the normalized state is thus given by 
\begin{eqnarray}
\label{psiev}
\fl |\psi(t+dt)\ra= \cases{|\psi(t)\ra + dt \left( -\rmi H_{S}-\frac{\gamma}{2}\pi_{1}  + \frac{1}{2} \alpha_t \right) |\psi(t)\ra,~ 
{\rm with~prob.}~p_0=1-\alpha_t dt, \label{matin}\\
|\psi_1\ra,~{\rm with~prob}~p_1=\alpha_t dt, \\
} \\
{\rm where}~ \alpha_t \equiv \gamma \la \psi(t)|\pi_1|\psi(t)\ra = \gamma|b(t)|^2.  \label{rate}
\end{eqnarray}
Equivalently, we can write also the complex non linear stochastic equation
\begin{equation}
\fl d\left|\psi(t)\right\rangle =\left(-iH_{S}-\frac{\gamma}{2}\pi_{1}+\frac{\alpha_{t}}{2}\right)\left|\psi(t)\right\rangle dt+\left(\sqrt{\gamma}\frac{\pi_{1}}{\sqrt{\alpha_{t}}}-I\right)\left|\psi(t)\right\rangle dN_{t}.
\label{Samedibeau}
\end{equation}
Note that for the second outcome in Eq.~(\ref{matin}) we should include a factor $b(t)/|b(t)|$. However, rigorously,  we should interpret the equation Eq.~ (\ref{Samedibeau}) (and all the others of this articles) for the corresponding one-point projector $|\psi (t)\ra \la \psi(t)|$.  
In more explicit vectorial way 
\begin{equation}
\fl d\left(\begin{array}{c}
a(t)\\
b(t)
\end{array}\right)=\left[\left(\begin{array}{cc}
\frac{\gamma}{2}\left|b(t)\right|^{2} & -i\gamma_{0}\\
-i\gamma_{0} & -\frac{\gamma}{2}+\frac{\gamma}{2}\left|b(t)\right|^{2}
\end{array}\right)dt+\left(\begin{array}{cc}
-1 & 0\\
0 & -1+\frac{1}{\left|b(t)\right|}
\end{array}\right)dN_{t}\right]\left(\begin{array}{c}
a(t)\\
b(t)
\end{array}\right).
\label{Samedibeau2}
\end{equation}

In these equations,  $N_{t}$ is a Poisson counting process which counts the  number of clicks  in any finite interval $[0,t]$. For almost all realizations one may take $N_{0}=0$. The change $dN_{t}$ at the instance $t$ be defined as the Ito-differential with the usual properties 
\begin{equation}
dN_{t}=N_{t}-N_{t-},\qquad dN_{t}\,dN_{t}=dN_{t},\qquad dN_{t}\,dt=0,
\label{dim}
\end{equation}
where $N_{t-}=\lim_{t'\rightarrow t-}N_{t'}$ (here we assume the trajectories of $N_t$ are right continuous with left limits) and with the expected value of the Poisson increment, conditioned upon the fact that the state of the ket of system
takes the value $|\psi(t)\ra$, is equal to 
\begin{equation}
\label{eq:RateF}
\mathbb{E}\left[dN_{t}\right]=\alpha_{t} dt = \gamma|b(t)|^2 dt.
\label{Samedi}
\end{equation}

Equations (\ref{Samedibeau}, \ref{Samedibeau2}) are sometimes called stochastic Schr\"odinger equations~\cite{wiseman_milburn_2009} or quantum trajectory for pure state. The first appearance of this type of equation with Poisson noise in this set-up was in \cite{DIOSI1986451}  and in \cite{PhysRevLett.68.580}. Since then,
different justifications have been given for the fact that they model quantum systems which
are subject to continuous indirect measurements.  The general case of quantum trajectories is for mixed states and includes also Gaussian white noise~\cite{Jacobs2006,wiseman_milburn_2009,breuer2002theory}.

For blind measurements one considers  an average over the outcomes and the density matrix $\rho(t)=\la |\psi(t)\ra \la \psi(t)| \ra$ evolves via
\begin{equation}
    \partial_t \rho(t)=-i[H_S, \rho(t)]+\frac{\gamma}{2} \left( 2 \pi_1 \rho (t) \pi_1 - \{\pi_1,\rho(t)\} \right).
\end{equation}
which is the form of the Lindblad equation~\cite{Lindblad1976,gorini1976} with only one Krauth operator $\pi_1$ which is moreover self-adjoint.

Physically, two phenomena are in competition in equations  (\ref{Samedibeau}, \ref{Samedibeau2}) : 
\begin{enumerate}
\item Collapsing in basis $|\psi_{0} \ra,|\psi_{1} \ra$  thanks to continuous measurement.     More precisely, when  $\gamma_0 \rightarrow0$ (i.e. $H_S \rightarrow0$), (\ref{Samedibeau}, \ref{Samedibeau2}) models the continuous measurement of $\pi_{1}$. As $\pi_{1}$ is a diagonal matrix in the basis $|\psi_{0} \ra,|\psi_{1} \ra$, this basis is said to be of non-demolition form with respect to the measurement~\cite{wiseman_milburn_2009}.  This will lead at large time to ~\cite{PhysRevA.84.044103,Adler_2001}  the collapse in the basis $|\psi_{0} \ra,|\psi_{1} \ra$ with the born law with respect to the initial ket $\left|\psi(0)\right\rangle$ , i.e. : 
\begin{eqnarray}
\lim_{t\rightarrow\infty}\left|\psi(t)\right\rangle =
\cases{
\left|\psi_{0}\right\rangle \textrm{ with probability } \left|\left\langle \psi_{0}\right|\left.\psi(0)\right\rangle \right|^{2}
\\
\left|\psi_{1}\right\rangle \textrm{ with probability } \left|\left\langle \psi_{1}\right|\left.\psi(0)\right\rangle \right|^{2}
}.
\label{collapse}
\end{eqnarray}
\item Rabi (coherent) oscillation due to unitary evolution. More precisely, when $\gamma = 0$, then (\ref{dim},\ref{Samedi}) $dN_t=0$ and  the equation (\ref{Samedibeau}, \ref{Samedibeau2}) is the  free (unitary) evolution  with Rabi Hamiltonian $H_{S}$ which leads to classical Rabi oscillation  \cite{le_bellac_2006}.  
\end{enumerate}
Note that in the general case with finite $\gamma_0$ and  $\gamma$, as the commutator $[H_{S},\pi_{1}]\neq0$, the unitary evolution comes to prevent the asymptotic collapse (\ref{collapse}). The asymptotic behavior will then be a smooth invariant density that we will exhibit below. Note also that the competition between continuous non demolition measurement and thermalization (instead of free unitary evolution here) has recently  been extensively studied (see e.g Refs:~\cite{Bauer_2015,Tilloy,Benoist2021}).

\subsection{No Click (deterministic) dynamics and Survival Probability : Saddle-node bifurcation in measurement parameter }
\label{sec:NoClick}
A continuous measurement of the object system corresponds to the case when $\tau\rightarrow 0$. In this limit, the un-normalized state $\tilde{\psi}(t)$ (Eq.~(\ref{eq:Noclick})) evolves via a non-Hermitian Hamiltonian when no clicks are registered. This evolution equation is
\begin{equation}
\label{eq:NoClickNonHerm}
\rmi \frac{\partial |\widetilde{\psi}\ra}{\partial t}=\gamma_{0} H_{\mathit{eff}} \, |\widetilde{\psi}\ra,\qquad H_{\mathit{eff}}=H_s-i\frac{\gamma}{2 \gamma_0} \pi_1=\left[{\begin{array}{cc}
0 & 1\\
1 & -2 \rmi \lambda
\end{array}}\right]
\end{equation}
where $\lambda=\frac{\gamma}{4\gamma_{0}}$ is to be regarded as a measurement parameter. The probability of no click being registered in the interval $[0,t]$ is \cite{dbd}
\begin{equation}
\label{eq:DefSu}
S(t)=\la\tilde{\psi}(t)|\tilde{\psi}(t)\ra.
\label{eq:dalida}
\end{equation}
Eq.~(\ref{eq:NoClickNonHerm}) is solved by matrix exponentiation in \ref{sec:CalSurvProb}. We shown in \ref{sec:CalSurvProb} that for $\lambda=1$, $H_{\mathit{eff}}$ is not a diagonalizable operator, otherwise it is. For the initial condition $|\widetilde{\psi}(0)\ra=|\psi(0)\ra=|\psi_{0}\ra$, one obtains from relations Eqs(\ref{eq:danser},\ref{eq:danser2}) the following expressions for survival probability for various values of the measurement parameter $\lambda$.

\begin{equation}
\label{eq:SurvProb}
S(t,\lambda) = \cases{
\case{\rme^{-\case{\gamma}{2}t}}{\beta^{2}}\left(\sin^{2}\left(\beta\gamma_{0}t\right)+\sin^{2}\left(\beta\gamma_{0}t+\phi\right)\right) &  for $0\leq \lambda < 1$ \\
\case{\rme^{-\case{\gamma}{2}t}}{\beta'^{2}}\left(\sinh^{2}\left(\beta'\gamma_{0}t\right)+\sinh^{2}\left(\beta'\gamma_{0}t+\phi'\right)\right) & for $\lambda > 1$ \\
\rme^{-\case{\gamma}{2}t}\left((\gamma_{0}t)^{2}+(1+\gamma_{0}t)^{2}\right) & for $\lambda = 1$\\}
\label{beau}
\end{equation}

In the above equation, one has $\beta^{2}=-\beta'^{2}=1-\lambda^{2}$ ,  $\sin\phi=\beta$ and $\sinh\phi'=\beta'$. Because $\beta'<\lambda$, even for $\lambda>1$ the survival probability is a decaying exponential. In all cases $\left[S(t,\lambda)\right]_{t\rightarrow\infty}=0$. The value $\lambda=1$ is clearly a crossover point where the form of the functional dependence of $S(t)$ changes. Furthermore, for a fixed $\gamma_{0}$ and $\lambda\ne 1$ one has,
\begin{equation}
\label{eq:alpha2}
\lim_{t\rightarrow\infty}\frac{S(t,1)}{S(t,\lambda)}=0.
\end{equation}
Thus the survival probability decays at the fastest rate for the critical value of $\lambda=1$.

The normalized state $|\psi(t)\ra$ follows a non linear equation when conditioned to evolve via no clicks. Noting that $|\psi(t)\ra=|\tilde{\psi}(t)\ra/\sqrt{S(t,\lambda)}$, after differentiation and use of Eq.~(\ref{eq:NoClickNonHerm}), one has
\begin{equation*}
\rmi \partial_t|{\psi}\ra=\gamma_{0} H_{\mathit{eff}}\, |\psi\ra-\frac{\rmi}{2}\left(\frac{d}{dt}\log S(t)\right)|\psi\ra.
\end{equation*}
 From Eqs.~(\ref{eq:NoClickNonHerm},\ref{eq:dalida}) and the above, it follows that
\begin{equation*}
-\rmi\frac{d}{dt}\log S(t)=\gamma_{0} \la\psi|\left[H_{\mathit{eff}}^{\dagger}-H_{\mathit{eff}}\right]|\psi\ra.
\end{equation*}
Combining the above two equations and Eq.~(\ref{eq:NoClickNonHerm}), one has the evolution equation for the normalized state $|\psi(t)\ra$
\begin{equation}
\label{eq:NoClickNonLin}
\rmi\partial_t{|\psi\ra}=\gamma_{0} H_{\mathit{eff}}\, |\psi\ra+\rmi 2 \lambda \gamma_{0}|b(t)|^2 |\psi \ra.
\label{se}
\end{equation}
Alternatively, this equation results also directly by taking $dN_t=0$ in Eqs.~(\ref{Samedibeau}, \ref{Samedibeau2}).
\begin{figure}
    \centering
    \includegraphics[scale=0.75]{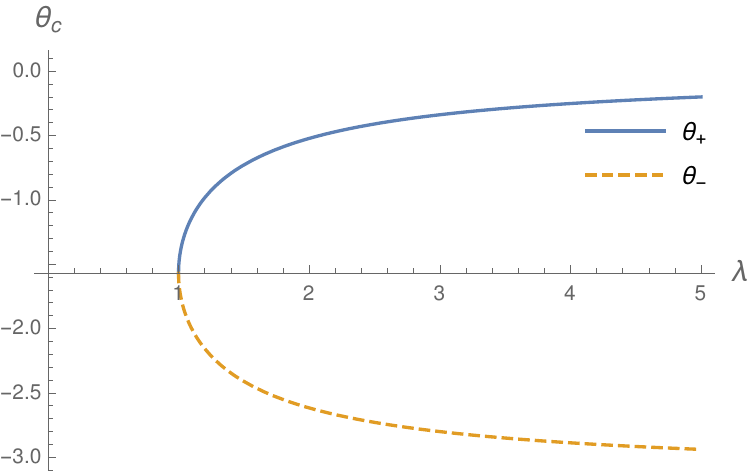}
    \caption{The saddle node bifurcation occurs at $\lambda=1$ for $\theta_{c}=-\pi/2$. For $\lambda>1$, two fixed points $\theta_{\pm}$ develop.}
    \label{fig:bifurc}
\end{figure}
\begin{figure}
	\centering
	\includegraphics[scale=0.85]{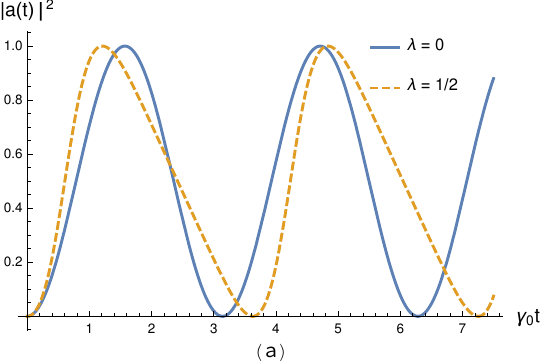}
  	\includegraphics[scale=0.85]{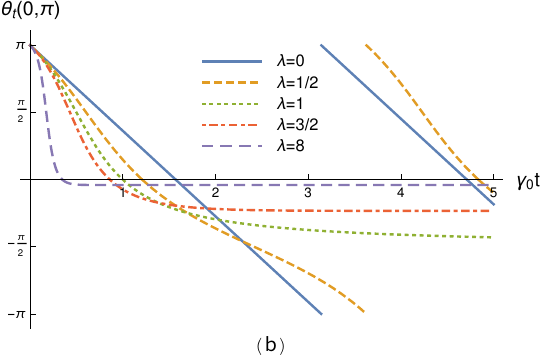}
     \caption{\emph{No click dynamics}: Figure (a) compares the oscillations of  the probablity $|a(t)|^{2}=\cos ^{2} \left(\theta_{t}(0,\pi)/2\right)$ for no measurement ($\lambda=0$) and with measurement  ($\lambda=1/2$). The early half of the cycle is covered faster than the later half for $\lambda=1/2$. (b) is the plot of $\theta_t$ for  $\lambda$ in various regimes. We chose initial condition $\theta_0=\pi$ but other choices would give qualitatively the same results.}
     \label{fig:DetEvol}
\end{figure}

{\bf Bloch sphere representation}: The pure state of a qubit can be represented by a point on the surface of the Bloch sphere whose northpole is the state $|\psi_0\rangle$ and the southpole is $|\psi_1\rangle$. It is  shown in \ref{sec:QubitEvol} that, for the particular choice of $H_{\mathcal{S}}$  and the starting initial conditions, $|\psi(t=0)\ra=|\psi_0\ra$,  $|\psi(t=0)\ra=|\psi_1\ra$ or point on the $yz$ plane, the qubit state remains in a fixed plane for all times  and  we can  use the following representation:
\begin{equation}
\label{eq:GenStateYZ}
|\psi(t)\ra=\left[{\begin{array}{c}
\cos \left(\theta_{t}/2\right)\\
\rmi \sin\left(\theta_{t}/2\right)
\end{array}}\right],
\end{equation}
where $\theta\in(-\pi,\pi]$ with $\pi$ and $-\pi$ are identified. Substituting this form of $\psi(t)$ in Eq(\ref{eq:NoClickNonLin}) gives the evolution equation for $\theta_{t}$ under no click dynamics
\begin{equation}
\label{eq:AngularSpeed}
\dot{\theta}= \Omega(\theta)=-2\gamma_{0} \left[1+\lambda\sin\theta\right].
\label{simple}
\end{equation}
This corresponds to the overdamped dynamics of a particle in a periodic tilted potential. For $\lambda <1  $
there are no fixed points and the particle keeps going round. At $\lambda=1$, there is a saddle-node bifurcation (Figure \ref{fig:bifurc}) and two fixed points develop, one of which is stable ($\theta_+$) and the other unstable ($\theta_-$) and given by:
\begin{equation}
\label{tpm}
 \theta_+=-\sin^{-1}(1/\lambda),\qquad \theta_-=-\pi-\theta_+.  
\end{equation}

All this is confirmed by  explicit integration of the above, see \ref{sec:integTheta}. In the following,  we use the flow notation $\theta_{t}(s,\theta')$ to indicate the solution $\theta_{t}$ of the no-click dynamics  $d\theta_{t}=\Omega(\theta_{t})\,dt$ such that at the instance $s$, $\theta_{s}=\theta'$. For $0<\lambda<1$, the probability $|a(t)|^{2}=\cos ^{2} \left(\theta_{t}(0,\pi)/2\right)$ shows oscillations similar to Rabi oscillations, which happen for $\lambda=0$. The frequency of these oscillations is proportional to $\beta$, whose form is given in  Eq.~(\ref{eq:alphaLT2}). Figure \ref{fig:DetEvol}(a)  compares the cases $\lambda=0$ and $\lambda=1/2$. Figure \ref{fig:DetEvol}(b) shows the evolution of $\theta_{t}(0,\pi)$ obtained from the integration of Eq.~(\ref{eq:AngularSpeed}) for various values of $\lambda$. We observe that the oscillatory behaviour stops exactly at $\lambda=1$ and increasing values of $\lambda \gg 1$  (collapsing regime of the previous section)  cause rapid decay to $\theta_{+}\approx 0$ which is consistent with the collapse  Eq.~(\ref{collapse}) but conditioned on no click.

\subsection{Stochastic  dynamics}
\label{sec:StochEval}
The continuous evolution of $\theta$ in accordance with Eq.~(\ref{eq:AngularSpeed}) is interrupted whenever a click occurs. In accordance with Eq.~(\ref{eq:collapse}), the system collapses to $\psi_{1}$ and hence the value of $\theta$ jumps to $\pi$  (we note that the azimuthal angle as defined in \ref{sec:QubitEvol} is undefined for this state). 

The full dynamics of $\theta_{t}$ consists of continuous evolution with occasional reset to $\theta=\pi$ whenever there is a click. This stochastic dynamics is described by 
\begin{equation}
\label{eq:SDEMain}
d\theta_{t}=\Omega\left(\theta_{t}\right)dt+\left(\pi-\theta_{t_{-}}\right)dN_{t},
\end{equation}
where the rate function $\alpha_{t}$ of $N_t$ depends on $\theta_{t}$ in accordance with (\ref{Samedi})
\begin{equation}
\label{eq:RateF}
\mathbb{E}\left[dN_{t}\right]=\alpha(\theta_t)\,dt=\gamma\sin^2\frac{\theta_t}{2}\,dt.
\end{equation}
Equivalently, we can write the dynamics in Eq.~\ref{eq:SDEMain} as
\begin{eqnarray}
\label{eq:SDEMain2}
\theta_{t+dt} = \cases{\theta_t+\Omega(\theta_{t})\,dt ~~~~~{\rm with ~prob.}~1-\alpha(\theta_t) dt \\
\pi \hskip 2.6cm{\rm with~ prob.}~ \alpha(\theta_t) dt}.
\end{eqnarray}
\begin{figure}
	\centering
	\includegraphics[scale=0.6]{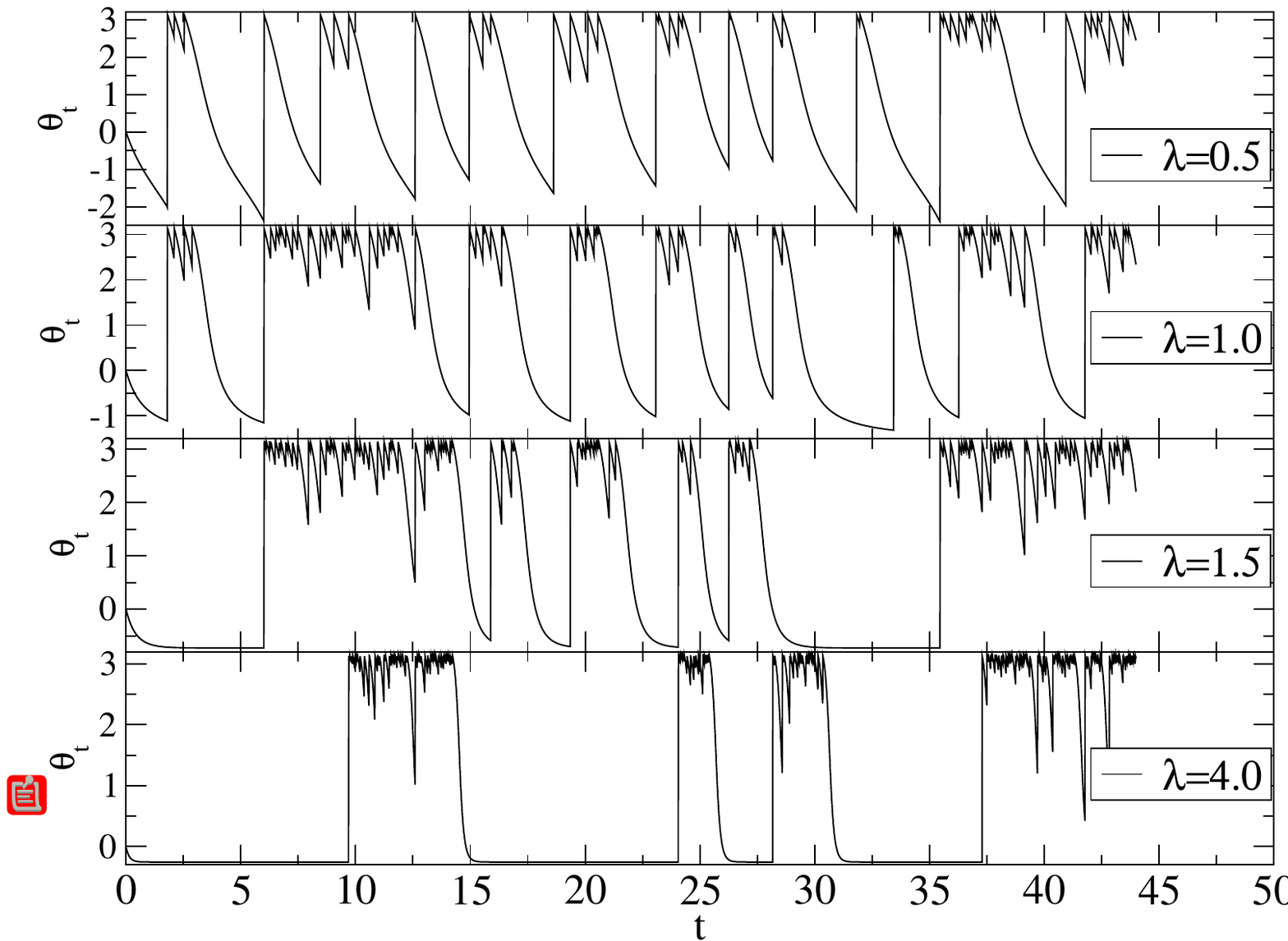}
        \caption{ Typical realizations of stochastic trajectories, $\theta_t$, obtained by solving Eq.~(\ref{eq:SDEMain2}) for 
        the intial condition $\theta_0=0$ and four different values of $\lambda$. We can see deterministic drifts and the stochastic resets to $\theta=\pi$.  
        For trajectories starting from $\theta_{0}=0$, the whole interval $[-\pi,\pi]$ is accessible for $\lambda=1/2$, while for $\lambda=3/2$ (more generally for all $\lambda\ge 1$), only the interval $[\theta_{+},\pi]$ is accessible. For the definition of $\theta_{+}$, see Eq.~(\ref{tpm}).}
        \label{fig:StochTr}
\end{figure}
In Fig.~(\ref{fig:StochTr}) we show typical stochastic trajectories obtained by evolving with this equation for different values of the strength, $\lambda$, of the resetting  rate.

Let $P(\theta,t)$ represent the probability density for $\theta$ at time $t$. It   satisfies the following master  equation~\cite{Snizhko}
\begin{equation}
\label{eq:PIDE}
\fl\frac{\partial P(\theta,t)}{\partial t}= -\frac{\partial }{\partial \theta}[\Omega(\theta)P(\theta,t)]-\gamma\sin^2\left(\frac{\theta}{2}\right)P(\theta,t)+\gamma\delta(\theta-\pi)\int_{0}^{2\pi}\sin^2\left(\frac{\theta'}{2}\right)P(\theta',t)\,d\theta'.~
\end{equation}
The steady state solution  of the above equation and some properties of the linear evolution operator were obtained in Ref.~\cite{Snizhko}. It was noted that the onset of Zeno dynamics occurs in several stages which are marked by specific values of the measurement parameter $\lambda\in\{1,\frac{2}{\sqrt{3}},2\}$. In Eq.~(\ref{eq:alpha2}), one already notices that the transition value $\lambda=1$ makes itself apparent in the rate of decay of the survival probability. In the following, we wish to investigate how these values of $\lambda$ appear in the counting statistics of the process $N_{t}$. We also provide a complete solution of the time-evolution of $P(\theta,t)$ and a more detailed characterization of the spectrum.

\section{Counting Statistics}
\label{sec:CountStat}

A stochastic process is in general described in terms of its finite dimensional probability densities. Given an initial state $\psi(0)$ of the object system, for the counting process $N_{t}$ these densities (exclusive probability densities, EPD) are \cite{Davies1969,Barchielli1991}
\begin{equation}
P_{0}^{t}[0|||\psi(0)\rangle],\, \, \, \, p_{0}^{t}[t_{1},\dots,t_{n}|||\psi(0)\rangle].
\label{eq:fdist}
\end{equation}
$P_{0}^{t}[0|||\psi(0)\rangle]$ is the probability of obtaining no clicks in the interval $ (0,t] $. $p_{0}^{t}\left[t_{1},\dots,t_{n}|| |\psi(0)\rangle\right]$ is the probability density (in times $t_{1},\dots,t_{n}$) of exactly n counts at instances $0<t_{1}<\dots<t_{n}\le t$ where $n$ ranges over positive integers. 

One has for the non-autonomous counting process $N_{t}$~\cite{Bremaud1981}
\begin{equation}
P_{0}^{t}[0|| \psi(0)]=\exp\bigg[-\int_{0}^{t}\alpha(\theta_{s}(0,\theta_0))\,ds\bigg].
\label{TA}
\end{equation}
For the $2-$state system under consideration,  $|\psi(0)\rangle=|\psi_{0}\rangle$ which corresponds to $\theta_{0}=0$ and $P_{0}^{t}[0|| |\psi(0)\rangle]$ is then nothing but the survival probability in Eq.~(\ref{eq:SurvProb}). 

From Eqs.~(\ref{eq:AngularSpeed}, \ref{eq:RateF}) we obtain
\begin{equation}
\int_{0}^{t}\alpha(\theta_{s}(0,\theta_{0}))\,ds=\frac{\gamma t}{2}+\log\left|\frac{1+\lambda\sin\theta_{t}(0,\theta_{0})}{1+\lambda\sin\theta_{0}}\right|.
\end{equation}
The survival probability can now be compactly written in the form 
\begin{equation}
\label{eq:AltSurvP}
P_{0}^{t}[0||\theta_{0}]=\frac{\Omega(\theta_{0})}{\Omega(\theta_{t}(0,\theta_{0}))}\,\rme^{-\frac{\gamma t}{2}}.
\label{iter}
\end{equation}
Using results from \ref{sec:integTheta} and Eq.~(\ref{eq:AltSurvP}), one recovers, in the case $\theta_0=0$, all the expressions in Eq.~(\ref{eq:SurvProb}).

Now consider the probability density in time $p_{0}^{t}[t_{1}|| \theta_{0}=0]$ of exactly one click at the instance $t_{1} \in (0,t]$. For this, there should be no click in $(0,t_{1}]$, a click in the interval $(t_{1},t_{1}+\Delta t_{1}]$ and no click from $(t_{1}+\Delta t_{1}, t]$. Then, in the limit $\Delta t\rightarrow 0$ one has  
\begin{equation}
\label{eq:OneClickD}
p_{0}^{t}[t_{1}|| \theta_{0}=0]= \rme^{-\frac{\gamma t}{2}}\frac{\Omega(0)}{\Omega(\theta_{t_{1}-}(0,0))}\times\alpha(\theta_{t_{1}-}(0,0))\times\frac{\Omega(\pi)}{\Omega(\theta_{t}(t_{1},\pi))}.
\end{equation}
Since $\theta_{0}=0$ in all further considerations, denote densities such as $p_{0}^{t}[t_{1}|| \theta_{0}=0]$ simply as $p_{0}^{t}[t_{1}]$ etc. For all $n\ge 1$, the densities $p_{0}^{t}[t_{1},\dots,t_{n}]$ can be obtained in a similar manner. For different values of $\lambda$, one may note the form of $p_{0}^{t}[t_{1},\dots,t_{n}]$ 
\begin{equation}
\label{eq:NClickDensity}
\fl p_{0}^{t}[t_{1},\dots,t_{n}]= \cases{
\case{\rme^{-\case{\gamma t}{2}}}{\beta^{2}}\left(\case{\gamma}{\beta^{2}}\right)^{n}\sin^{2}\left(\beta\gamma_{0} \,\Delta t_{0}\right)\frac{\prod_{k=1}^{n} \sin^{2}\left(\beta\gamma_{0}\,\Delta t_{k}-\phi\right)}{\sin^2(\theta_{t}(t_{n},\pi)/2)}   & $0\leq\lambda < 1$ \\
\case{\rme^{-\case{\gamma t}{2}}}{\beta'^{2}}\left(\case{\gamma}{\beta'^{2}}\right)^{n} \sinh^{2}\left(\beta'\gamma_{0}\,\Delta t_{0}\right)\frac{\prod_{k=1}^{n} \sinh^{2}\left(\beta\gamma_{0}\,\Delta t_{k}-\phi'\right)}{\sin^2(\theta_{t}(t_{n},\pi)/2)} & $\lambda > 1$ \\ 
\rme^{-\case{\gamma t}{2}} \gamma^{n}(\gamma_{0}\,\Delta t_{0})^2\frac{\prod_{k=1}^{n} (1-\gamma_{0}\,\Delta t_{k})^2}{\sin^2(\theta_{t}(t_{n},\pi)/2)} & $\lambda = 1$\\}
\end{equation}
 In the above equation, we have $\Delta t_{k} = t_{k+1}-t_{k}$ with $t_{0}=0$ and $t_{n+1} = t$. The expressions for $\sin^2(\theta_{t}(t_{n},\pi)/2)$ in the respective cases can be obtained from Eqs.~(\ref{eq:alphaLTz},  \ref{eq:alphaGT2z}, \ref{eq:alphaEQ2z}).
%

The probability of registering exactly $n$ counts in the interval $(0,t]$ is given by
\begin{equation}
\label{eq:nCountProb}
P_{0}^{t}[n]=\int_{0}^{t}dt_{n}\int_{0}^{t_{n}}dt_{n-1}\dots\int_{0}^{t_{2}}dt_{1}\,p_{0}^{t}[t_{1},\dots,t_{n}].
\end{equation}
Eq.~(\ref{eq:nCountProb}) allows for writing the moment generating function of $N_{t}$. Explicitly
\begin{equation}
\label{eq:MGF}
\mathbb{E}[\rme^{-s N_{t}}]=\sum_{n\ge 0}\rme^{-n s}P_{0}^{t}[n].
\end{equation}

In \ref{sec:MGFCalc}, it is shown that the Laplace transform with respect to time $t$ of the moment generating function is
\begin{equation}
\label{eq:LTransMGF}
\fl (\mathfrak{L}\mathbb{E}[\rme^{-s N_{t}}])(\sigma,s)=\frac{\mu^2-\frac{\gamma}{2}\mu+4\gamma_{0}^2}{\mu\left(\mu^2+4\beta^2\gamma_{0}^2\right)-\gamma \rme^{-s}\left(\mu^2-\frac{\gamma}{2}\mu+2\gamma_{0}^2\right)},
\end{equation}
where $\mu=\sigma+\gamma/2$. The denominator in the above is a third order polynomial in $\sigma$ and has in general three (possibly complex) zeros  $\sigma_{1},\,\sigma_{2},\,\sigma_{3}$ which depend on $s,\gamma,\gamma_{0}$. When these are distinct, then the moment generating function has the form
\begin{equation}
\label{eq:MGFcubic}
\fl \mathbb{E}[\rme^{-s N_{t}}]=
\frac{f(\sigma_{1}) \rme^{\sigma_{1}t} }{(\sigma_{1}-\sigma_{2})(\sigma_{1}-\sigma_{3})} + \frac{f(\sigma_{2})  \rme^{\sigma_{2}t} }{(\sigma_{2}-\sigma_{3})(\sigma_{2}-\sigma_{1})} + \frac{f(\sigma_{3}) \rme^{\sigma_{3}t}}{(\sigma_{3}-\sigma_{1})(\sigma_{3}-\sigma_{2})} ,
\end{equation}
where $f(\sigma_{i})$ is the numerator in Eq.~(\ref{eq:LTransMGF}) evaluated at the zero $\sigma_{i}$. In order to study the zeros, notice that the denominator factors when $s=0$ as 
\begin{equation}
\label{eq:DenomLaplaceT}
(\mu-2\lambda\gamma_{0})(\mu^2-2\lambda\gamma_{0}\mu+4\gamma_{0}^2).
\end{equation}
The zeros of the denominator in Eq.~(\ref{eq:LTransMGF}) evaluated at $s=0$ are therefore
\begin{equation}
\label{eq:zeros}
\sigma_{1}(0)=0,\,\,\, \sigma_{2}(0)=\gamma_{0}\left[-\lambda+\sqrt{\lambda^{2}-4}\right],\,\,\,
\sigma_{3}(0)=\gamma_{0}\left[-\lambda-\sqrt{\lambda^2-4}\right].
\end{equation}
The zeros are all real for $\lambda>2$. For $\lambda<2$, $\sigma_{2}(0)$ and $\sigma_{3}(0)$ are complex conjugate while for $\lambda=2$, there is a double root. When the expression in (\ref{eq:DenomLaplaceT}) is differentiated w.r.t. s and equated to $0$, then the following are obtained
\begin{equation}
\label{eq:Dzeros}
\frac{d \sigma_{1}}{ds}\bigg\vert_{s=0}=-\frac{\gamma}{2},\,\, \frac{d \sigma_{2}}{ds}\bigg\vert_{s=0}=\frac{\lambda}{\sqrt{\lambda^2-4}}\sigma_{3}(0),\\
\frac{d \sigma_{3}}{ds}\bigg\vert_{s=0}=-\frac{\lambda}{\sqrt{\lambda^2-4}}\sigma_{2}(0).
\end{equation}
For $\mathbb{E}[N_{t}]$ one has 
\begin{equation}
\label{eq:MeanCount}
\mathbb{E}[N_{t}]=-\frac{d}{ds}\mathbb{E}[\rme^{-s N_{t}}]\bigg\vert_{s=0}.
\end{equation}
A calculation using Eqs.~(\ref{eq:MGFcubic}, \ref{eq:zeros}, \ref{eq:Dzeros} \ref{eq:MeanCount}) then gives
\begin{equation}
\label{meanN}
\mathbb{E}[N_{t}]=\cases{
	2\lambda\gamma_{0}t+\lambda^{2}\left[-1+\rme^{-\lambda\gamma_{0}t}\,\frac{\sin(\omega t + \varphi)}{\sin \varphi}\right] & $0\le\lambda<2$, \\
	2\lambda\gamma_{0}t+\lambda^{2}\left[-1+\rme^{-\lambda\gamma_{0}t}\,\frac{\sinh(\omega' t + \varphi')}{\sinh \varphi'}\right] & $\lambda>2$, \\
	4\left(-1+\gamma_{0}t+\rme^{-2 \gamma_{0} t}(1+\gamma_{0}t)\right) & $\lambda=2$, \\}
\end{equation}
where $\omega^{2}=-\omega'^{2}=\gamma_{0}^{2}\left(4-\lambda^{2}\right)$, $\tan\varphi=\frac{\lambda\sqrt{4-\lambda^{2}}}{\lambda^{2}-2}$ and $\tanh\varphi'=\frac{\lambda\sqrt{\lambda^{2}-4}}{\lambda^{2}-2}$. Here again
one notices that $\lambda=2$ is a crossover point where the form of functional dependence of $\mathbb{E}[N_{t}]$ changes. Furthermore, for the value of $\lambda=\sqrt{2}$, the oscillatory function $\sin(\omega t+\varphi)/\sin(\varphi)$ is of minimum amplitude. In ~\cite{Snizhko}, it has been pointed that there exists a transition  for $\lambda=2/\sqrt{3}$ characterised by a divergence in the steady state probability density $P_{\infty}(\theta)(=\lim_{t\rightarrow{\infty}} P(\theta,t))$.  Moreover, the mean value of the transition rate $\alpha$ given by 
\begin{equation}
\label{eq:ExpIntensity1}
\overline{\alpha}_{t}=\frac{d}{dt}\mathbb{E}[N_{t}]=\gamma\Bigg[\frac{1}{2}-\frac{\gamma_{0}}{\omega}\rme^{-\lambda\gamma_{0}t}\left(\frac{\rme^{\rmi\omega t}}{\omega/\gamma_{0}+\rmi\lambda}+\frac{\rme^{-\rmi\omega t}}{\omega/\gamma_{0}-\rmi\lambda}\right)\Bigg]
\end{equation}
has the signature of $\lambda=2$ transition only.

It is important to note the limiting behaviour of expressions for $S(t,\lambda)$ in Eq.~(\ref{eq:SurvProb}) and $\mathbb{E}[N_{t}]$ in Eq.~(\ref{meanN}). As defined, $\lambda=\frac{\gamma}{4\gamma_{0}}$ and two possible ways for $\lambda\rightarrow\infty$ are that $\gamma_{0}\rightarrow 0$ for fixed $\gamma$, and that $\gamma\rightarrow \infty$ for fixed $\gamma_{0}$. In either case, it is easy to see that 
\begin{equation}
\label{eq:zeno}
    \lim_{\lambda\rightarrow \infty} S(t,\lambda)=1.
\end{equation}
As expected in the Zeno effect, for most trajectories the experimenter would detect no clicks in a finite time under strong measurement. For $\mathbb{E}[N_t]$, the behaviour is quite different, as we see in the following limits.
\begin{equation}
\label{eq:spike}
    \lim_{\lambda\rightarrow\infty \atop \gamma \,\mathit{fixed}} \mathbb{E}[N_t]=0,\quad\quad \lim_{\lambda\rightarrow\infty \atop \gamma_{0} \,\mathit{fixed}} \mathbb{E}[N_t]=2\gamma_{0}^{2}t^{2}.
\end{equation}
The second limit above is the more interesting one. For large $\gamma$ (with $\gamma_{0}$ fixed), the transition rate $\alpha$ (Eq.\ref{eq:RateF}) is of the order $\gamma$ for $\theta\approx\pi$. Immediately after the first click, $\theta=\pi$ and with a high click rate, the experimenter is likely to observe a large number of subsequent clicks. This is also evident in the numerical simulation as can be seen in the last panel  in Fig.~(\ref{fig:StochTr}).  Thus for large $\gamma$, durations of no clicks (darkeness) are punctuated by durations of a rapid increase in the number of clicks (brightness)\cite{Cohen-Tannoudji_1986}. In the limit of large $\gamma$, the second limit in Eq.~(\ref{eq:spike}) is achievable despite Eq.~(\ref{eq:zeno}). This can be related to  the phenomenon studied in the context of spiking and collapse in the large noise limit of stochastic differential equations driven by Wiener processes \cite{Clement10.1214/22-AAP1819}. We contend that the spikes have signature in the noise statistics of the Poisson signal in the model we consider.

\section{Exact time-dependent solution for $P(\theta,t)$}
\label{sec:SteadyState}

\subsection{Formal solution from renewal approach} For the  resetting process, one could write a solution to Eq.~(\ref{eq:PIDE}) directly with the aid of the EPDs mentioned in (\ref{eq:fdist}). A given value of $\theta$ can be attained at time $t$ after no reset, after exactly $1$ reset, after exactly $2$ resets and so on. These are all mutually exclusive events. Summing their contributions, one has
\begin{equation*}
\fl P(\theta,t) = P_{0}^{t}[0] \delta(\theta-\theta_{t}(0,0))+\sum_{n\ge 1}\int_{0}^{t}\dots\int_{0}^{t_{2}}\, p_{0}^{t}[t_{1},\dots,t_{n}]\delta(\theta-\theta_{t}(0,0)) \prod_{k=1}^{n}dt_{k}.
\end{equation*}
In the $n^{\mathit{th}}$ summand, suppose the last reset occurred at $t_{n} = t-\tau$. Then from the property in Eq.~(\ref{eq:OneClickD}), one has 
\begin{equation}
\eqalign{
\fl p_{0}^{t}[t_{1},\dots,t_{n}]\delta(\theta-\theta_{t}(0,0)) &= p_{0}^{t-\tau}[t_{1},\dots,t_{n-1}]\,\gamma\sin^{2}\left(\case{\theta_{(t-\tau)-}(0,0)}{2}\right)\,P_{t-\tau}^{t}[0||\pi]\,\delta(\theta-\theta_t(t-{\tau},{\pi})) \nonumber\\
&=p_{0}^{t-\tau}[t_{1},\dots,t_{n-1}]\,\underbrace{\gamma\sin^{2}\left(\case{\theta_{(t-\tau)-}(0,0)}{2}\right)}_{\alpha_{t-\tau}}\,P_{0}^{\tau}[0||\pi]\,\delta(\theta-\theta_{\tau}(0,{\pi})).}
\end{equation}
After substitution, one arrives at the formal solution which is in the form of a renewal equation~\footnote{We note that these type of renewal equations  have been discussed in the context of stochastic resetting in Refs.~\cite{pal2016,RoldanPhysRevE.96.022130}}:
\begin{equation}
\label{eq:FPESol}
P(\theta,t) = P_{0}^{t}[0] \delta(\theta-\theta_{t}(0,0))+\int_{0}^{t}\overline{\alpha}_{t-\tau} P_{0}^{\tau}[0||\pi]\,\delta(\theta-\theta_{\tau}(0,{\pi}))\,d\tau,
\end{equation}
where $\overline{\alpha}_{t-\tau}$ is the mean transition rate that has already been obtained in Eq.~(\ref{eq:ExpIntensity1}).  Alternatively, with the definition
\begin{equation}
\label{eq:ExpIntensity2}
\overline{\alpha}_{t}=\int_{0}^{2\pi}\gamma\sin^{2}\left(\frac{\theta}{2}\right)\,P(\theta,t)\,d\theta,
\end{equation}
when Eq.~(\ref{eq:FPESol}) is multiplied throughout by $\sin^{2}{\theta}/{2}$, integrated w.r.t. $\theta$ and the Laplace transform is taken, one obtains
\begin{equation}
[\mathfrak{L}\overline{\alpha}](\sigma)=\frac{\gamma}{\beta^{2}}\frac{\hat{g}_{0}}{1-\frac{\gamma}{\beta^{2}}\hat{g}_{\phi}}. \label{alpLT}
\end{equation}
in the notation of Eq.~(\ref{eq:LTransOfg}). Upon inversion, we recover Eq.~(\ref{eq:ExpIntensity1}).  One can now obtain the explicit form for $P(\theta,t)$ from Eqs.~(\ref{eq:ExpIntensity1},\ref{eq:FPESol}) in the various regimes of $\lambda$. 
\vspace{1cm}

\subsection{Steady state}
The evaluation of the steady state density $P_{\infty}(\theta)=\lim_{t\rightarrow\infty}P(\theta,t)$ is particularly simple. Since the time-dependent part of $\overline{\alpha}_t$ as well as the $P_{0}^{t}[0] \delta(\theta-\theta_{t}(0,0))$ contribution to $P(\theta,t)$ are exponentially suppressed, one has
\begin{equation}
\label{eq:SteadyState}
P_{\infty}(\theta)=\frac{\gamma}{2}\int_{0}^{\infty}P_{0}^{\tau}[0||\pi]\,\delta(\theta-\theta_{\tau}(0,{\pi}))\,d\tau.
\end{equation}

Consider the case $\lambda<1$. For $\theta_{0}=\pi$, the no-click evolution happens via
\begin{equation}
\label{eq:lambdaLTone}
\tan\frac{\theta_{\tau}(0,\pi)}{2}=
-\frac{\sin(\beta\gamma_{0}\tau-\phi)}{\sin(\beta\gamma_{0}\tau)},\,\,\,P_{0}^{\tau}[0||\pi]=\frac{\rme^{-\frac{\gamma\tau}{2}}}{1+\lambda\sin\theta_{\tau}(0,\pi)}.
\end{equation}
The same value of $\theta_{\tau}(0,\pi) = \theta$ modulo $2\pi$ can be attained at the times $\{\tau_{n}\}_{n\ge 0}$ where $\tau_{n}=\tau_{0}+\frac{n\pi}{\beta\gamma_{0}}$. The value of $\tau_{0}$ can be worked out to be
\begin{equation*}
\frac{\gamma\tau_{0}(\theta)}{2}=\frac{2\lambda}{\sqrt{1-\lambda^{2}}}\Big[\frac{\pi}{2}-\tan^{-1}\left(\frac{\lambda+\tan\frac{\theta}{2}}{\sqrt{1-\lambda^{2}}}\right)\Big].
\end{equation*}
In this case one also has
\begin{equation*}
\delta(\theta-\theta_{\tau}(0,\pi))=\sum_{n\ge 0}\frac{\delta(\tau-\tau_{n})}{|\Omega(\theta_{\tau_{n}}(0,\pi))|}=\frac{1}{2\gamma_{0}(1+\lambda\sin\theta)}\sum_{n\ge 0}\delta(\tau-\tau_{n}).
\end{equation*}
Integrating Eq.~(\ref{eq:SteadyState}) with the above information one obtains for measurement parameter $\lambda<1$
\begin{equation}
\label{eq:SSlambdaLT1}
P_{\infty}(\theta)=\frac{\lambda}{(1+\lambda\sin\theta)^{2}}\frac{\rme^{-\frac{\gamma\tau_{0}}{2}}}{1-\rme^{-\frac{2\pi\lambda}{\sqrt{1-\lambda^{2}}}}}.
\end{equation}
The case $\lambda\ge 1$, can be handled similarly. The main difference from $\lambda<1$ case is that there exists only $1$ instance $\tau_{0}$ when a given value $\theta$ can be attained as long as $\theta$ does not lie in the no-go region. For $\lambda=1$, one has
\begin{equation}
\label{eq:SSlambdaEQ1}
\frac{\gamma\tau_{0}(\theta)}{2}=\frac{2}{1+\tan\frac{\theta}{2}},\qquad P_{\infty}(\theta)=\frac{\rme^{-\frac{\gamma\tau_{0}}{2}}}{(1+\sin\theta)^{2}} 1_{(-\frac{\pi}{2},\pi]}(\theta).
\end{equation}
For $\lambda>1$, one has
\begin{equation}
\label{eq:SSTauLambdaGT1}
\eqalign{ \rme^{-\frac{\gamma\tau_{0}(\theta)}{2}}=\left(\frac{\tan\frac{\theta}{2}-\tan\frac{\theta_{+}}{2}}{\tan\frac{\theta}{2}-\tan\frac{\theta_{-}}{2}}\right)^{\frac{\lambda}{\sqrt{\lambda^2-1}}},\qquad\tan\frac{\theta_{\pm}}{2}=-\lambda\pm\sqrt{\lambda^2-1}.\\
P_{\infty}(\theta)=\frac{\lambda \rme^{-\frac{\gamma\tau_{0}}{2}}}{(1+\lambda\sin\theta)^{2}} 1_{(\theta_{+},\pi]}(\theta).}
\end{equation}
\begin{figure}
	\centering
	\includegraphics[scale=0.26]{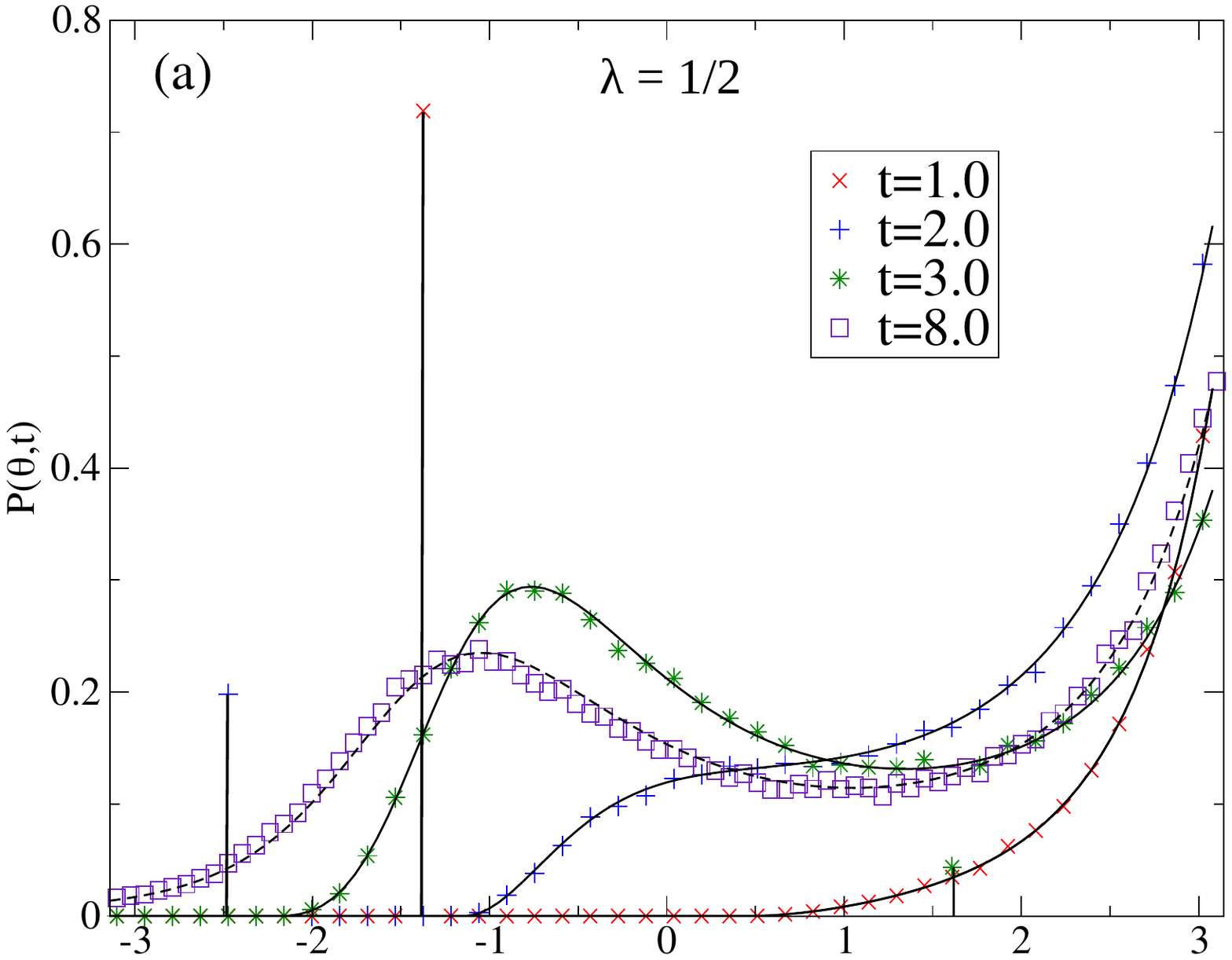}
	\includegraphics[scale=0.26]{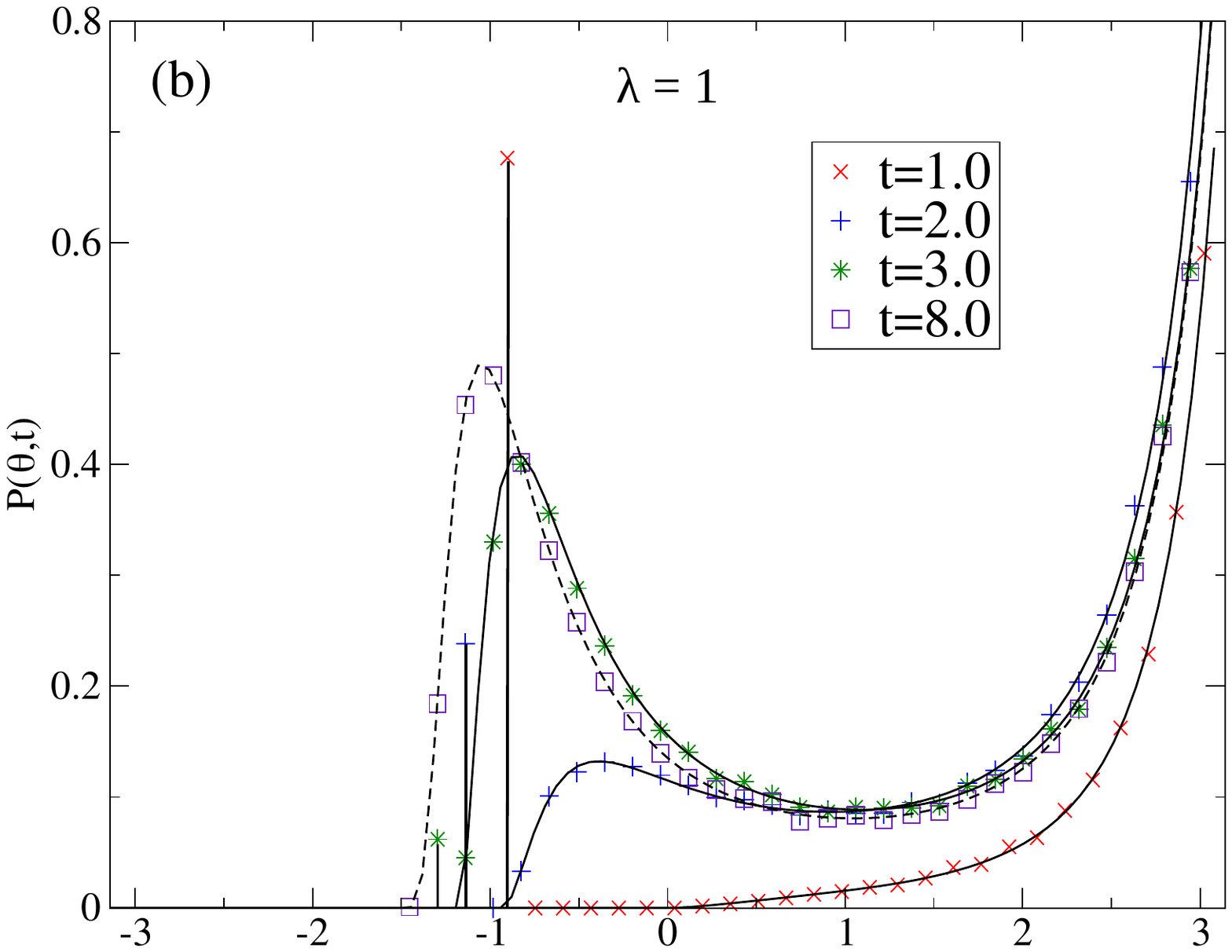}
	\begin{minipage}[c]{0.45\textwidth}
		\includegraphics[scale=0.26]{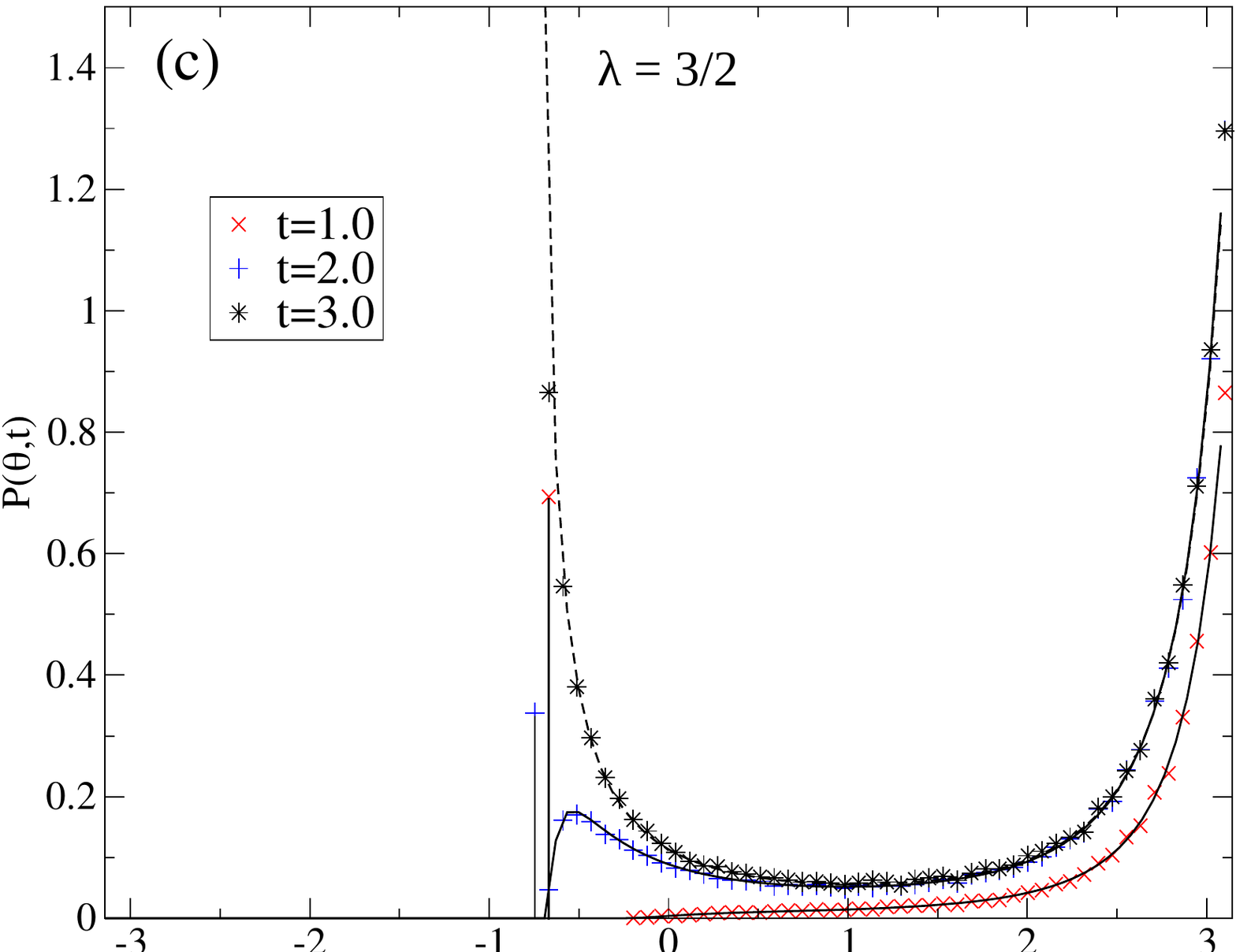}
  \end{minipage}   
\begin{minipage}[c]{0.54\textwidth}
	\caption{Time evolution of $P(\theta,t)$ is shown for (a) for $\lambda=1/2$, (b) for  $\lambda=1$ and (c) for $\lambda=3/2$. The black solid lines  are from the analytic solution from  Eq.~(\ref{eq:FullSolLambdaLT1}) in (a), Eq.~(\ref{eq:FullSolLambdaEQ1}) in (b) and Eq.~\ref{eq:FullSolLambdaGT1} in (c).  The points were generated from simulations of $10^5$ trajectories. The dashed black lines are the analytic results for the steady state $P_{\infty}(\theta)$ from Eqs.~(\ref{eq:SSlambdaLT1},\ref{eq:SSlambdaEQ1},\ref{eq:SSTauLambdaGT1}). The spikes correspond to the $\delta$ function propagating with the no-click dynamics and the height of the peaks equals the probability mass on the $\delta$ function.}
        \label{fig:Comparisons}
    \end{minipage}
\end{figure}
The results contained in Eqs.~(\ref{eq:SSlambdaLT1},\ref{eq:SSlambdaEQ1},\ref{eq:SSTauLambdaGT1}) agree with those obtained in Ref.~\cite{Snizhko} by directly finding the steady state solution of  Eq.~(\ref{eq:PIDE}). With the resetting approach and through use of the renewal equation,  we are now able to obtain the explicit time dependence of $P(\theta,t)$ in all cases. We  remark that in the Zeno limit of strong measurement $\lambda \gg 1$  (coming from $\gamma$ finite and $\gamma_0 \rightarrow 0$), $\theta_{+}\rightarrow 0$ and we expect the steady state to converge towards a singular density concentrated near $0$ and $\pi$.


\subsection{Time evolution}
The following equations note the result of integration of Eq.~(\ref{eq:FPESol}) for finite time. The integration can be carried out by use of Eqs.~(\ref{eq:AltSurvP},\ref{eq:ExpIntensity1}) and the flow equations in \ref{sec:integTheta} for the respective cases of $\lambda$.

For $\lambda<1$, the formula is somewhat complicated because the indicator function $1_{(\theta_{t}(0,\pi),\pi]}$ has to wrap properly with the number of possible visits to $\theta$ in time $t$ starting from $\theta_{0}=\pi$. For $\gamma_{0} t\le \pi/\beta$, there is only one possible visit and the expression is simpler. We give this expression:
\begin{equation}
\label{eq:FullSolLambdaLT1}
\eqalign{
\fl P(\theta,t)=P_{0}^{t}[0] \delta(\theta-\theta_{t}(0,0))+\frac{\lambda \rme^{-\frac{\gamma\tau_{0}}{2}}}{(1+\lambda\sin\theta)^{2}}\,1_{(\theta_{t}(0,\pi),\pi]}(\theta)\\
-\frac{4\lambda}{\sqrt{4-\lambda^{2}}}\frac{1_{(\theta_{t}(0,\pi),\pi]}(\theta)}{(1+\lambda\sin\theta)^{2}}{\rm Re }\left[\rme^{(-\lambda+\rmi\sqrt{4-\lambda^{2}})\gamma_{0}t}\frac{\rme^{-(\lambda+\rmi\sqrt{4-\lambda^{2}})\gamma_{0}\tau_{0}}}{\omega/\gamma_{0}+\rmi\lambda}\right]. \\}
\end{equation}
For $\lambda=1$, we get for all times $t>0$:
\begin{equation}
\eqalign{
\label{eq:FullSolLambdaEQ1}
\fl P(\theta,t)=P_{0}^{t}[0] \delta(\theta-\theta_{t}(0,0))\\
+\Bigg[1-\frac{2 \rme^{-\gamma_{0}(t-\tau_{0})}}{\sqrt{3}}\sin\left(\sqrt{3}\gamma_{0}(t-\tau_{0})+\frac{\pi}{3}\right)\Bigg]\frac{\rme^{-\frac{\gamma\tau_{0}}{2}}}{(1+\sin\theta)^{2}} 1_{(\theta_{t}(0,\pi),\pi]}(\theta),&}
\end{equation}
while for $\lambda>1$, we get (for all times $t>0$):
\begin{equation}
\label{eq:FullSolLambdaGT1}
\eqalign{\fl P(\theta,t)=P_{0}^{t}[0] \delta(\theta-\theta_{t}(0,0))\\
+\Bigg[1-\frac{2 \rme^{-\lambda\gamma_{0}(t-\tau_{0})}}{\sqrt{4-\lambda^{2}}}\sin\left(\omega(t-\tau_{0})+\arctan\sqrt{\frac{4}{\lambda^{2}}-1}\right)\Bigg]\frac{\lambda \rme^{-\frac{\gamma\tau_{0}}{2}}}{(1+\lambda\sin\theta)^{2}} 1_{(\theta_{t}(0,\pi),\pi]}(\theta).}
\end{equation}  
The three graphs  in Figure \ref{fig:Comparisons} show good agreement between simulation and the analytic forms in Eqs.~(\ref{eq:FullSolLambdaLT1},\ref{eq:FullSolLambdaEQ1} \& \ref{eq:FullSolLambdaGT1}).

\subsection{General formulation and explicit time Laplace transform solution for the transition probability}
We consider here a more general set-up where the transition probability
$P_t(\theta|\theta')$, that the process pass from $\theta'$ at time
$0$ to $\theta$ at time $t$,  solves the Kolomogorov equation 
\begin{eqnarray}
\fl \partial_{t}P_t(\theta|\theta')=L\left(\theta\right) P_t(\theta|\theta')- \gamma(\theta) P_t(\theta|\theta') +\mu(\theta)\left(\int_{0}^{2\pi}d\theta''\gamma(\theta'')P_t(\theta''|\theta')\right).\nonumber \\
\end{eqnarray}
Here, $L\left(\theta\right)$ is a second (resp. first) order differential operator in $\theta$, markov
generator,  coming from a diffusion (resp. deterministic) process and we are considering that resetting is not to a particular point but to a point $\theta$ chosen from the probability distribution $\mu(\theta)$ and the positive function $ \gamma(\theta)$ is the jump rate for escape from state $\theta$. The
associated master equation for the probability density $P(\theta,t)$
is then 
\begin{eqnarray}
\fl \partial_{t}P(\theta,t)=L\left(\theta\right) P(\theta,t)- \gamma(\theta) P(\theta,t)+\mu(\theta)\left(\int_{0}^{2\pi}d\theta''\gamma(\theta'')P(\theta'',t)\right).
\end{eqnarray}
Let us  define  the operators 
\begin{eqnarray}
L_{0}\left[f\right]\left(\theta\right)\equiv L\left(\theta\right)\left[f\right]\left(\theta\right)-\left\langle \mu,1\right\rangle\gamma(\theta)f(\theta), \\
L_{1}\left[f\right]\left(\theta\right)\equiv\mu(\theta)\left\langle \gamma,f\right\rangle,
\label{eq:L-L}
\end{eqnarray}
for any function $f$ on $\left[0,2\pi\right]$ and where have used the following inner product definition:
\[
\left\langle f,g\right\rangle  = \int_{0}^{2\pi}d\theta f(\theta) g(\theta).
\]
The master equation (\ref{eq:PIDE}), and more generally the set-up of the previous section,  is a particular case of this general theory when $L\left(\theta\right)[f]=-\partial_\theta [ \Omega\left(\theta\right) f(\theta)]$,  $\gamma\left(\theta\right)=\gamma\sin^{2}\left(\frac{\theta}{2}\right)$
and $\mu\left(\theta\right)=\delta\left(\theta-\pi\right).$ 
With these definitions, we have
the formal solution of the Kolmogorov equation $P_{t}=\exp\left(t\left(L_{0}+L_{1}\right)\right)$ and the time Laplace transform is the resolvent \begin{eqnarray}
\fl \left(\mathcal{\mathfrak{L}}P_{t}\right)(s)&\equiv\int_{0}^{\infty}dt\exp\left(-st\right)\exp\left(t\left(L_{0}+L_{1}\right)\right)=\left(s-L_{0}-L_{1}\right)^{-1} \\
\fl &=\left(s-L_{0}\right)^{-1}+\left(s-L_{0}\right)^{-1}L_{1}\left(s-L_{0}-L_{1}\right)^{-1}.
\end{eqnarray}
We thus find the auto-consistency relation 
\[
\left(\mathcal{\mathfrak{L}}P_{t}\right)(s)=\left(\mathcal{\mathfrak{L}}P_{t}^{(0)}\right)(s)+\left(\mathcal{\mathfrak{L}}P_{t}^{(0)}\right)(s)L_{1}\left(\mathcal{\mathfrak{L}}P_{t}\right)(s),
\]
 where $P_{t}^{(0)}=\exp\left(tL_{0}\right)$.
By plugging in this relation the expression for $L_{1}$ (\ref{eq:L-L}) we obtain 
\begin{eqnarray}
\fl \left(\mathcal{\mathfrak{L}}P_{t}\right)(s)(\theta|\theta')=\left(\mathcal{\mathfrak{L}}P_{t}^{(0)}\right)(s) (\theta|\theta')+
 \left\langle\left(\mathfrak{L}P_{t}^{(0)}\right)(s) \left(\theta|.\right)  ,\mu (.) \right\rangle
\left\langle 
\gamma (.),\left(\mathcal{\mathfrak{L}}P_{t}\right)(s)(.|\theta')
\right\rangle,\nonumber \\ \label{Pgensola}
\end{eqnarray}
where $.$ indicates the  inner product integration variable. 
Multiplying with $\gamma(\theta)$ and integrating over $\theta$ solves for the unknown last term in the above equation
\begin{eqnarray}
\left\langle 
\gamma (.),\left(\mathcal{\mathfrak{L}}P_{t}\right)(s)(.|\theta') \right\rangle = \frac{\left\langle 
\gamma (.),\left(\mathcal{\mathfrak{L}}P^0_{t}\right)(s)(.|\theta') \right\rangle }{1-\left\langle \gamma(..), \left\langle\left(\mathfrak{L}P_{t}^{(0)}\right)(s) \left(..|.\right)  ,\mu (.) \right\rangle    \right\rangle}. \label{Pgensolb}
\end{eqnarray}
Equations (\ref{Pgensola}) and (\ref{Pgensolb}) provide a complete solution of the time evolution in the Laplace domain. We have finally the exact relation  
\begin{eqnarray}
\fl \left(\mathcal{\mathfrak{L}}P_{t}\right)(s)(\theta|\theta')
=
\left(\mathcal{\mathfrak{L}}P_{t}^{(0)}\right)(s) (\theta|\theta')
+
\frac{\left\langle\left(\mathfrak{L}P_{t}^{(0)}\right)(s) \left(\theta|.\right)  ,\mu (.) \right\rangle \left\langle 
\gamma (.),\left(\mathcal{\mathfrak{L}}P^0_{t}\right)(s)(.|\theta') \right\rangle }{1-\left\langle \gamma(..), \left\langle\left(\mathfrak{L}P_{t}^{(0)}\right)(s) \left(..|.\right)  ,\mu (.) \right\rangle    \right\rangle},\nonumber \\ \label{Pgensola}
\end{eqnarray}
which expresses  $\left(\mathcal{\mathfrak{L}}P_{t}\right)(s)$ in terms of $\left(\mathcal{\mathfrak{L}}P_{t}^{(0)}\right)(s)$. So, in the case where the second is explicit, so is the first. It is instructive to  write this equation in the time domain. To this end, we denote by   $\bar{\gamma}(t)$  the inverse Laplace transform of {the l.h.s of} Eq.~\ref{Pgensolb}, which is simply
\begin{eqnarray}
\bar{\gamma}_t(\theta') = \int_0^{2 \pi} d \theta'' ~\gamma(\theta'') P_t(\theta''|\theta'), 
\end{eqnarray}
which is just the average of transition rate {$\gamma(\theta_t)$ conditioned by the initial condition $\theta_0=\theta'$}. Then Eq.~\ref{Pgensola} in the time domain is given by
\begin{eqnarray}
P_t(\theta|\theta')=P^0_t(\theta|\theta')+ \int_0^t d\tau \langle P^0_\tau(\theta|.),\mu(.)\rangle \bar{\gamma}_{t-\tau}(\theta'),
\end{eqnarray}
which we see has the same structure as the renewal equation in Eq.~(\ref{eq:FPESol}). The   Eq.~(\ref{Pgensolb}) {gives the Laplace transform of $\bar{\gamma}_t$, and}  is then Eq.~(\ref{alpLT}), and for our specific example we were able to compute $\bar{\gamma}_t \equiv \bar{\alpha}_t$ explicitly (from the inverse Laplace and also using a renewal approach). In general, we would have an explicit solution for $P_t$ from Eq.~(\ref{Pgensola}), provided we are able to evaluate  $\bar{\gamma}_t$ explicitly from Eq.~(\ref{Pgensolb}).

\section{Spectral Analysis}
\label{sec:spectral}
The Fokker-Planck operator corresponding to  Eq.~(\ref{eq:PIDE}) has  interesting spectral  properties which were pointed out in \cite{Snizhko}. We wish to extend those studies and in particular, for the case $0< \lambda \leq 1$, we report some new results and some subtle features.  One seeks solutions to Eq.~(\ref{eq:PIDE}) in the form $P(\theta,t)=\exp(2\gamma_{0}\,\nu\,t)\,f_{\nu}(\theta)$. This leads to the following eigenvalue problem  for the operator $\mathcal{L}$:
\begin{equation}
\label{eq:Eval0}
\fl \mathcal{L} f_{\nu}= (1+\lambda\sin\theta)\partial_{\theta}f_{\nu}+\lambda\left(2 \cos\theta-1\right)\,f_{\nu}
 +2 \lambda\delta(\theta-\pi)\int_{0}^{2\pi}\sin^2(\theta'/2) f_\nu(\theta')\,d\theta'=\nu\,f_{\nu}.~~~
\end{equation}
Noting that $\pi$ and $-\pi$ are identified, we integrate the above equation over a small interval across $\pi$.  Assuming that $f_\nu(\theta)$ has no divergence at $\pi$ this then gives a discontinuity of $f_\nu$ across $\pi$ and we are led  to the following equivalent set of equations:
\begin{eqnarray}
\mathcal{L} f_{\nu}=(1+\lambda\sin\theta)\partial_{\theta}f_{\nu}+\lambda\left(2 \cos\theta-1\right)\,f_{\nu}=\nu\,f_{\nu}, \label{eq:EValProbA} \\
f_\nu(\pi-0)-f_{\nu}(\pi+0)=2\lambda\int_{-\pi}^{\pi}\sin^{2}\left(\theta/2\right)\,f_{\nu}(\theta)\,d\theta.\label{eq:EValProbB}
\end{eqnarray}
We also define  an operator $\mathcal{L}_{0}$ which satisfies Eq.~(\ref{eq:EValProbA}) but with boundary conditions   satisfying $f(\pi-0)=f(\pi+0)$.
The adjoint operator $\mathcal{L}_{0}^{\dagger}$ acts on square integrable functions $g(\theta)$ as
\begin{equation}
\label{eq:EValProb0Adj}
\mathcal{L}_{0}^{\dagger}g = -\left(1+\lambda\sin\theta\right)\partial_{\theta}g-2\lambda\sin^{2}\left(\theta/2\right)\,g,\quad g(\pi-0)=g(\pi+0).
\end{equation}
Since $\mathcal{L}_{0}\mathcal{L}_{0}^{\dagger}\ne\mathcal{L}_{0}^{\dagger}\mathcal{L}_{0}$, $\mathcal{L}_{0}$ is not a normal operator and it's eigenfunctions do not form an orthonormal basis for the Hilbert space $\mathbb{L}^{2}[-\pi,\pi]$. The same holds for the operator pair $\mathcal{L}$, $\mathcal{L}^{\dagger}$. Because of the integral boundary condition in Eq.~(\ref{eq:EValProbB}), one has
\begin{equation}
\label{eq:AdjMain}
\mathcal{L}^{\dagger}h = -\left(1+\lambda\sin\theta\right)\partial_{\theta}h-2\lambda\sin^{2}\left(\theta/2\right)\,(h-h(\pi)),\quad h(\pi-0)=h(\pi+0).~~
\end{equation}
We see below that the eigenfunctions of $\mathcal{L}_{0}$ and $\mathcal{L}_{0}^{\dagger}$ together form a bi-orthonormal basis.

\subsection{Measurement parameter $0\le\lambda<1$}
\label{subSec:LambdaLT1}

 If $\bar{f}_{\nu}$ is an eigenfunction  of $\mathcal{L}_{0}$ with eigenvalue $\nu$ then it is easily seen that
\begin{equation}
\label{eq:SOLlt1}
\fl \bar{f}_{\nu}(\theta)=\frac{C_{\nu}}{(1+\lambda\sin\theta)^{2}}\exp\Bigg[\frac{\nu+\lambda}{\sqrt{1-\lambda^{2}}}\,\varphi(\theta,\lambda)\Bigg],\qquad \varphi(\theta,\lambda)=2\arctan\left(\frac{\lambda+\tan\frac{\theta}{2}}{\sqrt{1-\lambda^{2}}}\right),
\end{equation}
where $C_{\nu}$ is a normalization constant chosen so that $\int_{-\pi}^{\pi}\bar{f}_{\nu}(\theta),d\theta=1$.
 The value of $\varphi(\theta,\lambda)$ are defined at the boundary by $\varphi(\pm\pi,\lambda)=\lim_{\theta\rightarrow\pm\pi^{\mp}}\varphi(\theta,\lambda)=\pm\pi$.
On imposing the boundary condition    $\bar{f}_{\nu_{m}}(\pi)=\bar{f}_{\nu_{m}}(-\pi_{+})$, the eigenvalues are obtained to be $\nu_{m}=-\lambda+\rmi m \sqrt{1-\lambda^{2}}$ where $m$ ranges over the set of integers. Similar calculations for the operator $\mathcal{L}_{0}^{\dagger}$ gives its spectrum. The following equation gives a complete bi-orthonormal system, of eigenfunctions and eigenvalues, for the pair $\mathcal{L}_{0}$, $\mathcal{L}_{0}^{\dagger}$ with the property $\la g_{\nu_{m}},\bar{f}_{\nu_{n}}\ra=\delta_{mn}$.
\begin{equation}
\label{eq:EFL0}
\eqalign{
\fl \bar{f}_{\nu_{m}}(\theta)=\left(\case{1-\lambda^{2}}{4\pi^{2}}\right)^{\case{1}{4}}\frac{\exp[\rmi m \varphi(\theta,\lambda)]}{(1+\lambda\sin\theta)^{2}},\qquad\nu_{m}=-\lambda+\rmi m \sqrt{1-\lambda^{2}},m\in\mathbb{Z}\qquad 
\mbox{for}\,\, \mathcal{L}_{0},\\
\fl g_{\nu_{m}}=\left(\case{1-\lambda^{2}}{4\pi^{2}}\right)^{\case{1}{4}}(1+\lambda\sin\theta)\exp[\rmi m \varphi(\theta,\lambda)],\quad\nu_{-m}=-\lambda-\rmi m \sqrt{1-\lambda^{2}},m\in\mathbb{Z}\quad 
\mbox{for}\,\, \mathcal{L}_{0}^{\dagger}.}
\end{equation}
One can  write a canonical expansion for  any function $f\in \mathbb{L}^{2}[-\pi,\pi]$ in terms of the basis $f_{\nu_{m}}$.
\begin{equation}
\label{eq:CanonicalExpFree}
f = \sum_{m\in\mathbb{Z}}\alpha_{m}\,f_{\nu_{m}},\qquad\alpha_{m}=\la g_{\mu_{m}},f\ra=\int_{-\pi}^{\pi}g_{\mu_{m}}^{*} f\,d\theta.
\end{equation}

For the eigenvalue problem in Eq.~(\ref{eq:EValProbA}), the functions in Eq.~(\ref{eq:SOLlt1}) still satisfy the formal equation but the boundary condition in Eq.~(\ref{eq:EValProbB}) leads to the condition
\begin{equation}
\label{eq:EVALSlt1}
\frac{\nu(\nu^{2}+\lambda\nu+1)}{(\nu+\lambda)(\nu^{2}+2\nu\lambda+1)}\sinh\Big[\frac{\nu+\lambda}{\sqrt{1-\lambda^{2}}}\pi\Big]=0.
\end{equation}
From above, we infer that the eigenvalues of $\mathcal{L}$  are $\{0,\nu_{+},\nu_{-},\nu_{m}\}$ for $m\in\mathbb{Z}\setminus\{-1,0,1\}$ where
\begin{equation}
\label{eq:BoundStateEvals}
\nu_{\pm}=[-\lambda\pm\rmi\sqrt{4-\lambda^{2}}]/2.
\end{equation}
The $\sinh$ function vanishes for $\nu=\nu_{m},\,\forall m\in\mathbb{Z}$, however the denominator itself is $(\nu-\nu_{0})(\nu-\nu_{1})(\nu-\nu_{-1})$. The limiting value of the ratio for these three choices of $\nu$ is non-zero and therfore  $\mathcal{L}$ does not have $\nu_{0}$, $\nu_{-1}$ and $\nu_{1}$ as eigenvalues. When the eigenvalue problem for adjoint $\mathcal{L}^{\dagger}$ in Eq.~(\ref{eq:AdjMain}) is solved, one obtains the same contraint in Eq.~(\ref{eq:EVALSlt1}). Thus the operators $\mathcal{L}$ and $\mathcal{L}^{\dagger}$ have the same set of eigenvalues just like the operators $\mathcal{L}_{0}$ and $\mathcal{L}_{0}^{\dagger}$. We further note that the integral boundary condition in Eq.~(\ref{eq:EValProbB}) displaces only three eigenvalues in spectrum of  the $\mathcal{L}_{0}$, $\mathcal{L}_{0}^{\dagger}$ system --- i.e  $f_{\nu_m}=\bar{f}_{\nu_m}$ for $m \neq 0,\pm 1$, while for these three eigenvalues $\{0,\nu_{+},\nu_{-}\}$ we obtain three different eigenstates:
\begin{equation}
\label{eq:DiscSpecL}
\fl f_{0}(\theta)=\frac{\lambda}{2\sinh\left[\case{\pi(\lambda)}{\sqrt{1-\lambda^{2}}}\right]}\frac{\exp\left[\frac{\lambda}{\sqrt{1-\lambda^{2}}}\varphi(\theta,\lambda)\right]}{(1+\lambda\sin\theta)^{2}},~~ f_{\nu_{\pm}}(\theta)=\frac{\nu_{\mp}}{2\sinh\left[\case{\pi\nu_{\mp}}{\sqrt{1-\lambda^{2}}}\right]}\frac{\exp\left[-\frac{\nu_{\mp}}{\sqrt{1-\lambda^{2}}}\varphi(\theta,\lambda)\right]}{(1+\lambda\sin\theta)^{2}} .~~
\end{equation}
We note that the $\nu=0$ eigenvector $f_0(\theta)$ corresponds to the steady state solution $P_\infty(\theta)$. 

For the eigenvalue problem in Eq.(\ref{eq:AdjMain}), with the convention $\mathcal{L}^{\dagger}h_{\nu^{*}}=\nu h_{\nu^{*}}$, where $\nu^*$ denotes the complex conjugation, the differential equation can be rewritten as
\begin{equation*}
\fl \frac{\partial}{\partial\theta}\left[\frac{h_{\nu^{*}}(\theta)-h_{\nu^{*}}(\pi)}{1+\lambda\sin\theta}\exp\left[\frac{\lambda+\nu}{\sqrt{1-\lambda^{2}}}\varphi(\theta,\lambda)\right]\right]=-\frac{\nu\,h_{\nu^{*}}(\pi)}{(1+\lambda\sin\theta)^{2}}\exp\left[\frac{\lambda+\nu}{\sqrt{1-\lambda^{2}}}\varphi(\theta,\lambda)\right].
\end{equation*}
Integrating the above from $\pi$ to $\theta$, one obtains, after changing the integration variable on the r.h.s  $\theta \to \varphi$  
\begin{eqnarray}
\fl \frac{h_{\nu^{*}}(\theta)-h_{\nu^{*}}(\pi)}{1+\lambda\sin\theta}\exp\left[\frac{\lambda+\nu}{\sqrt{1-\lambda^{2}}}\varphi(\theta,\lambda)\right]=\\ \nonumber
-\frac{\nu h_{\nu^{*}}(\pi)}{(1-\lambda^{2})^{3/2}}\int_\pi^{\varphi(\theta)} d \varphi \exp\left[\frac{\lambda+\nu}{\sqrt{1-\lambda^{2}}}\varphi\right] \left(1-\frac{\lambda}{2}\nu_{-1}\rme^{\rmi\varphi}-\frac{\lambda}{2}\nu_{1}\rme^{-\rmi\varphi}\right). \nonumber
\end{eqnarray}
After performing the integration and applying the boundary condition in Eq.~(\ref{eq:AdjMain}), we  obtain the  eigenfunctions  $\{h_{0},h_{\nu_{+}},h_{\nu_{-}},h_{\nu_{m}}\} $  ($m\in\mathbb{Z}\setminus\{-1,0,1\}$) for  $\mathcal{L}^{\dagger}$. They are indexed such that  $\mathcal{L}^{\dagger} h_{\nu}=\nu^{*}h_{\nu}$. The bi-orthonormality  condition $\la h_{a},f_{b}\ra=\delta_{ab}$ can be verified from their explicit form: 
\begin{equation}
\label{eq:lambdaLT1CompleteSet}
\eqalign{
h_{0}=1,~~
 h_{\nu_{\pm}}(\theta)=\mp\rmi\frac{\lambda \big[\cos\theta-\nu_{\mp}\sin\theta\big]}{\sqrt{4-\lambda^{2}}},\\
 h_{\nu_{m}}(\theta)=g_{\nu_{m}}(\theta)+\frac{(-1)^{m}\lambda^{2}} {B_{-m}}\sum_{k\in\{-1,0,1\}}\frac{m(m^{2}-1)}{(m-k)(k^{2}+1)}\frac{g_{\mu_{k}}(\theta)}{B_{k}},
}
\end{equation}
where  $g_{\mu_{m}}$ is defined in Eq.~(\ref{eq:EFL0}). The coefficients $B$ appearing in the above expressions are 
\begin{equation*}
B_{m}=\nu_{m}(\nu_{m}-\nu_{-})(\nu_{m}-\nu_{+}).
\end{equation*}
In this bi-orthonormal system, one can now expand
\begin{equation*}
\delta(\theta)=f_{0}+h_{\nu_{-}}(0)\,f_{\nu_{+}}+h_{\nu_{+}}(0)\,f_{\nu_{-}}+\,\sum_{m\in\mathbb{Z}\setminus\{-1,0,1\}}\,h_{\nu_{-m}}(0)\,f_{\nu_{m}}.
\end{equation*}
In the original problem Eq.~(\ref{eq:PIDE}), $P(\theta,0)=\delta(\theta)$. Then the time development of $P(\theta,t)$ is given by
\begin{eqnarray}
\fl P(\theta,t)= f_{0}(\theta)+\rmi\frac{\lambda\,\rme^{2\nu_{+}\gamma_{0}\,t}}{\sqrt{4-\lambda^{2}}}f_{\nu_{+}}(\theta) - \rmi\frac{\lambda\,\rme^{2\nu_{-}\gamma_{0}\,t}}{\sqrt{4-\lambda^{2}}}f_{\nu_{-}}(\theta)+\left(\frac{1-\lambda^{2}}{4\pi^{2}}\right)^{\frac{1}{4}}\sum_{m\in\mathbb{Z}\setminus\{-1,0,1\}}\frac{(-1)^{m}\lambda} {B_{m}}f_{\nu_{m}}(\theta)\rme^{2\nu_{m}\gamma_{0}\,t} \nonumber\\
\fl +\left(\frac{1-\lambda^{2}}{4\pi^{2}}\right)^{\frac{1}{2}}\frac{\rme^{-2\lambda\gamma_{0}\,t}}{(1+\lambda\sin\theta)^{2}}\sum_{m\in\mathbb{Z}\setminus\{-1,0,1\}} \exp\bigg[\rmi\,m\,\Phi(\theta,\lambda,t)\bigg], 
\end{eqnarray}
where $\Phi(\theta,\lambda,t)= \varphi(\theta,\lambda)-\varphi(0,\lambda)+2\gamma_{0}\,t\sqrt{1-\lambda^{2}}$.  We note that the first series in the rhs is convergent as $B_{m}\sim m^{3}$.
We write the last summation in the above equation as $ 2\pi \delta [\Phi(\theta,\lambda,t)]-\sum_{m\in\{-1,0,1\}} \exp\bigg[\rmi\,m\,\Phi(\theta,\lambda,t)\bigg]$. Then using the fact that  $\Phi(\theta,\lambda,t)=0$ solves for $\theta_{t}(0,0)$ (Eq.~(\ref{eq:alphaLT2})), we obtain $\delta [\Phi(\theta,\lambda,t)]= \delta(\theta-\theta_t(0,0))/|\Phi'(\theta,\lambda,t)|$. In conjunction with Eq.~(\ref{eq:AltSurvP}), after some simplifications, we finally get
\begin{equation}
\label{eq:SpectralLT1}
\eqalign{
\fl P(\theta,t)=P_{0}^{t}[0]\delta(\theta-\theta_{t}(0,0))+ P_{\mbox{f}}(\theta,t), ~~{\rm where}\\
\fl P_{\mbox{f}}(\theta,t) = P_\infty(\theta)+\rmi\frac{\lambda\,\rme^{2\nu_{+}\gamma_{0}\,t}}{\sqrt{4-\lambda^{2}}}f_{\nu_{+}}(\theta) - \rmi\frac{\lambda\,\rme^{2\nu_{-}\gamma_{0}\,t}}{\sqrt{4-\lambda^{2}}}f_{\nu_{-}}(\theta)+\left(\frac{1-\lambda^{2}}{4\pi^{2}}\right)^{\frac{1}{4}}\sum_{m\in\mathbb{Z}\setminus\{-1,0,1\}}\frac{(-1)^{m}\lambda} {B_{m}}f_{\nu_{m}}(\theta)\rme^{2\nu_{m}\gamma_{0}\,t}\\
-\left(\frac{1-\lambda^{2}}{4\pi^{2}}\right)^{\frac{1}{2}}\frac{\rme^{-2\lambda\gamma_{0}\,t}}{(1+\lambda\sin\theta)^{2}}\frac{\sin\left[3\Phi(\theta,\lambda,t)/2\right]}{\sin\left[\Phi(\theta,\lambda,t)/2\right]}.}
\end{equation}
Here $P_{\mbox{f}}$ represents the finite part of the density which was also obtained in Eq.~(\ref{eq:FullSolLambdaLT1}). We have numerically verified the agreement between  Eq.~(\ref{eq:SpectralLT1}) and Eq.~(\ref{eq:FullSolLambdaLT1}). In $P_{\mbox{f}}(\theta,t)$, the smallest decay rate is for the terms corresponding to $f_{\nu_{+}}$ and $f_{\nu_{-}}$. Therefore the approach to $P_\infty(\theta)$ (the steady state) happens as\begin{equation}
\fl \rmi\frac{\lambda\,\rme^{2\nu_{+}\gamma_{0}\,t}}{\sqrt{4-\lambda^{2}}}f_{\nu_{+}}(\theta) - \rmi\frac{\lambda\,\rme^{2\nu_{-}\gamma_{0}\,t}}{\sqrt{4-\lambda^{2}}}f_{\nu_{-}}(\theta)=\rmi \frac{\lambda \rme^{-\lambda\gamma_{0}\,t}}{\sqrt{4-\lambda^{2}}}\left(f_{\nu_{+}}(\theta)\rme^{\rmi\gamma_{0}\,t\sqrt{4-\lambda^{2}}}-f_{\nu_{-}}(\theta)\rme^{-\rmi\gamma_{0}\,t \sqrt{4-\lambda^{2}}}\right),
\end{equation}
which has the form of a damped oscillator of natural 
frequency $2\gamma_0$. {Note that the spectral solution for $P(\theta,t)$ provides an easier result for the long time form than that obtained from the renewal solution.}

\subsection{Measurement parameter $\lambda=1$}
\label{subSec:LambdaEQ1}
For the operators $\mathcal{L}_{0}$ and $\mathcal{L}_{0}^{\dagger}$ defined in Eq.~(\ref{eq:EValProb0Adj}), the corresponding eigenfunctions and eigenvalues are \footnote {We note that $f_{\nu_{k}}$ in Eq.~(\ref{eq:EFL0lambda1}) satisfies the condition $f_{\nu_{k}}(\pi)=f_{\nu_{k}}(-\pi_{+})$ for any complex $k$. Our choice of $\nu_{k}$ for $k\in\mathbb{R}$ is based on the fact that this set is the limit of the set $\{\nu_{m}\}_{m\in\mathbb{Z}}$ (see Eq.~(\ref{eq:EFL0})) as $\lambda\rightarrow 1_{-}$.}
\begin{equation}
\label{eq:EFL0lambda1}
\eqalign{
\fl f_{\nu_{k}}=\frac{1}{\sqrt{2\pi}}\frac{\exp\left[\rmi k \frac{-2}{1+\tan\case{\theta}{2}}\right]}{(1+\sin\theta)^{2}},\qquad\nu_{k}=-1+\rmi k,k\in\mathbb{R}\qquad 
\mbox{for}\,\, \mathcal{L}_{0},\\
\fl g_{\mu_{k}}=\frac{1}{\sqrt{2\pi}}(1+\sin\theta)\exp\left[\rmi k \frac{-2}{1+\tan\case{\theta}{2}}\right],\qquad\mu_{k}=-1-\rmi k,k\in\mathbb{R}\qquad 
\mbox{for}\,\, \mathcal{L}_{0}^{\dagger}.}
\end{equation}
One notices that the eigenvalues are no more discrete and therefore the bi-orthonormality condition becomes $\la g_{\mu_{k}},f_{\nu_{k'}}\ra=\delta(k-k')$ , which is easily verified. While the functions $f_{\nu_{k}}$ satisfy the condition $f_{\nu_{k}}(-\pi)=f_{\nu_{k}}(\pi)$, they do not belong to $\mathbb{L}^{2}[-\pi,\pi]$. With the substitution $x=-2/(1+\tan\frac{\theta}{2})$, one has
\begin{equation}
\label{eq:BCLambda1}
\fl \int_{-\pi}^{\pi}2 \sin^{2}\case{\theta}{2} \,f_{\nu_{k}}\,d\theta = \frac{1}{2\sqrt{2\pi}}\int_{-\infty}^{\infty}(x+2)^{2}\,\rme^{\rmi k x} \,dx= 2\sqrt{2\pi}\left[\delta(k)-\rmi \delta'(k)-\delta''(k)/4\right].
\end{equation}
We shall develop the solution $P(\theta,t)$ in the complete system of Eqs.(\ref{eq:EFL0lambda1}). For $\theta_{0}\ne -\pi/2$, consider the integral
\begin{eqnarray*}
\fl \int_{-\infty}^{\infty}g_{\mu_{k}}(\theta_{0})^{*}\,f_{\nu_{k}}(\theta)\,dk=\frac{1+\sin\theta_{0}}{(1+\sin\theta)^{2}}\frac{1}{2\pi}\int_{-\infty}^{\infty} \exp\left[\rmi k\left(\frac{2}{1+\tan\case{\theta_{0}}{2}}- \frac{2}{1+\tan\case{\theta}{2}}\right)\right]\,dk \\ \nonumber
=\left(\frac{1+\sin\theta_{0}}{1+\sin\theta}\right)^{2} \delta(\theta-\theta_{0}).
\end{eqnarray*}
Since the factor of $\delta(\theta-\theta_{0})$ is continuous at $\theta_{0}$, one has the following representations
\begin{equation}
\label{eq:DeltaReps}
\delta(\theta)=\frac{1}{\sqrt{2\pi}}\int_{-\infty}^{\infty}\,\rme^{\rmi 2 k }\,f_{\nu_{k}}\, dk,\qquad
\delta(\theta-\pi)=\frac{1}{\sqrt{2\pi}}\int_{-\infty}^{\infty}\,f_{\nu_{k}}\, dk .
\end{equation}
The general function $P(\theta,t)$ can be expanded in the bi-orthonormal system of Eq.~(\ref{eq:EFL0lambda1}) as
\begin{equation}
\label{eq:PExpLambda1}
P(\theta,t) = \int_{-\infty}^{\infty} c_{k}(t) \,\rme^{2\nu_{k}\gamma_{0}t}\, f_{\nu_{k}}\,dk,
\end{equation}
where the time development of $c_{k}$ can be ascertained after substituting the above in Eq.~(\ref{eq:PIDE}). Doing so, using the Eqs.(\ref{eq:BCLambda1},\ref{eq:DeltaReps}) and the fact that $\mathcal{L}_{0}f_{\nu_{k}} = \nu_{k} f_{\nu_{k}}$, one has
\begin{equation}
\label{eq:CoeffDE}
\fl \dot{c}_{k} = \dot{c}_{0} \, \rme^{-\rmi\,k2\gamma_{0}t},\qquad \dot{c}_{0}=4\gamma_{0}\left[\left((1-\gamma_{0}t)+\frac{\rmi}{2}\frac{\partial}{\partial k}\right)^{2}\,c_{k}\right]_{k=0},\qquad c_{k}(0)=\frac{\rme^{2 \rmi k}}{\sqrt{2\pi}}.
\end{equation}
This is a self consistent system which can be readily solved using Laplace transform. We note the explicit solution
\begin{eqnarray}
\label{eq:CoeffSol}
\fl c_{k}(t)= \frac{1}{\sqrt{2\pi}}\Bigg[\rme^{\rmi 2 k}+\rmi \frac{\exp[-\rmi 2 \gamma_{0}t(k+\rmi)]-1}{k+\rmi}+\frac{\nu_{-}}{\sqrt{3}}\frac{\exp[-\rmi 2 \gamma_{0}t(k-\rmi \nu_{-})]-1}{k-\rmi \nu_{-}} \nonumber\\
-\frac{\nu_{+}}{\sqrt{3}}\frac{\exp[-\rmi 2 \gamma_{0}t(k-\rmi \nu_{+})]-1}{k-\rmi \nu_{+}}\Bigg],
\end{eqnarray}
where $\nu_{\pm}$ are as defined in Eq.~(\ref{eq:BoundStateEvals}) for $\lambda=1$. Substituting for $c_{k}$ in Eq.~(\ref{eq:PExpLambda1}) and carrying out  contour integration, $P(\theta,t)$ is obtained in the form of Eq.~(\ref{eq:FullSolLambdaEQ1}).

The point spectrum of $\mathcal{L}$ consists of the eigenvalues $\{0,\nu_{+},\nu_{-}\}$ with eigenfunctions
\begin{equation}
\fl f_{0}(\theta)=\frac{\exp\left[-\frac{2}{1+\tan(\theta/2)}\right]}{(1+\sin\theta)^{2}}1_{[-\pi/2,\pi]}(\theta),\qquad f_{\nu_{\pm}}(\theta)=-\nu_{\mp} \frac{\exp\left[\frac{2\nu_{\mp}}{1+\tan(\theta/2)}\right]}{(1+\sin\theta)^{2}}1_{[-\pi/2,\pi]}(\theta).
\end{equation}
These functions properly belong to $\mathbb{L}^{2}[-\pi,\pi]$ and satisfy the integral boundary condition in Eq.~(\ref{eq:EValProbB}). The continuous spectrum of $\mathcal{L}$ consists of improper eigenvalues $\nu_{k}$ with corresponding improper eigenfunction $f_{\nu_{k}}$ as defined in Eq.~(\ref{eq:EFL0lambda1}) for $k\in\mathbb{R}\setminus\{0\}$. These functions satisfy the boundary condition only upto principal value as is evident from Eq.~(\ref{eq:BCLambda1}).

Now consider the operator $\mathcal{L}^{\dagger}$ defined in Eq.~(\ref{eq:AdjMain}) for the case $\lambda=1$. An integration of the eigenvalue equation for eigenfunction $h_{\nu^{*}}$ of eigenvalue $\nu$ leads to the expression
\begin{eqnarray*}
\fl \frac{h_{\nu^{*}}(\theta)-h_{\nu^{*}}(\pi)}{1+\sin\theta} \exp\left[-(\nu+1)\frac{2}{1+\tan\frac{\theta}{2}}\right]=\frac{\nu h_{\nu^{*}}(\pi)}{(\nu+1)^{3}}\Bigg[1+\nu+\nu^{2} \\
-\frac{\left(2+2\nu+\nu^{2}+(1+\nu)\cos\theta+\sin\theta\right)}{1+\sin\theta}\exp\left[-(\nu+1)\frac{2}{1+\tan\frac{\theta}{2}}\right]\Bigg].
\end{eqnarray*}
In case of $\nu=-1$, the rhs of the above should by evaluated as a limit. This limit is 
\begin{equation*}
-h_{\nu^{*}}(\pi)\frac{2 \sec ^2\left(\theta/2\right) (-3 \sin\theta +\cos\theta -5)}{3 \left(\tan \left(\theta /2\right)+1\right)^3}.
\end{equation*}
Imposing the boundary condition, one obtains the discrete part of the spectrum to be $\{0,\nu_{+},\nu_{-}\}$ (same as for $\mathcal{L}$) with corresponding eigenfunctions given by $\{h_{0}, h_{\nu_{-}},h_{\nu_{+}}\}$ of the same form as in Eq.~(\ref{eq:lambdaLT1CompleteSet}). For the continuous part of the spectrum, we take $\nu_{k}$ as given in Eq.~(\ref{eq:EFL0lambda1}). The following gives the explicit form of $h_{\nu_{k}}$ which are proper eigenfunctions of  $\mathcal{L}^{\dagger}$ with eigenvalues $-1-\rmi k,\,\,k\in\mathbb{R}\setminus\{0\}$.
\begin{equation}
\fl h_{\nu_{k}} = g_{\mu_{k}}+\frac{1}{\sqrt{2\pi}}\left[\frac{1}{\nu_{-k}}-\frac{(\nu_{-k}+1)\cos\theta+\sin\theta}{\nu_{-k}^{2}+\nu_{-k}+1}\right],\qquad \mathcal{L}^{\dagger}h_{\nu_{k}}=\nu_{-k}h_{\nu_{k}}.
\end{equation}
The improper eigenfunction with improper eigenvalue $-1$ is give by 
\begin{equation}
h_{\nu_{0}}=\frac{1}{\sqrt{2\pi}}\left[1+\frac{2}{3}\left(\frac{3\sin\theta-\cos\theta +5}{1+\tan(\theta/2)}\right)\right].
\end{equation}

\section{Conclusion}
\label{sec:Conc}
We studied the dynamics of a qubit that is  continuously monitored via measurements on a detector qubit with which it interacts strongly so as to avoid the zeno limit. For the special choice of system Hamiltonian and initial conditions that we considered here, the  qubit state remains confined at all times on the $yz$ plane of the Bloch sphere so that it can represented by a single angle variable.  The state $|\psi(t) \ra $ follows a stochastic dynamics with  drift and jump  terms. We obtained various results for this dynamics. We summarize here our main findings:
\begin{itemize}
    \item We point out that the  stochastic  wavefunction dynamics can be naturally interpreted as a resetting process, with a resetting rate that depends on the instantaneous state. The strength of the resetting rate $\lambda$ quantifies the strength of  measurements. 
    \item We obtain exact results on the number of resetting events, $N_t$, in a specified time $t$. We show that the form of the time-dependence, of the mean number of events Eqs.~\ref{meanN}, has a transition at $\lambda=2$. 
    \item Using two different approaches, a renewal approach and one based on non-perturbative resolvent (or Green's function) approach, we obtain the exact form of the probability distribution $P(\theta,t)$ for the system to be in the quantum state,  $|\theta\ra=\left[{\begin{array}{c}
\cos \left(\theta/2\right)\\
\rmi \sin\left(\theta/2\right)
\end{array}}\right]$, at time $t$. At long times we recover the steady state form known from earlier studies. We show that as for the steady state, the time evolution has three different forms for the regimes $0 \leq \lambda <1$, $\lambda=1$ and $\lambda >1$. 
    \item For the cases $0 \leq \lambda <1$ and  $\lambda=1$ we evaluate the complete spectrum of the Fokker-Planck operator which forms a bi-orthonormal set. This provides another solution for the time evolution of $P(\theta,t)$, that is especially useful at long times.
    A future study will explore the more difficult  case $\lambda >1$.
\end{itemize}
We note that the average density matrix of the qubit is given by $\hat{\rho}(t) = \int_0^{2 \pi} d \theta P(\theta,t) |\theta \ra \la\theta|$. However, this density matrix contains much less information about the system. For example, in the steady state we have $\hat{\rho}(t) \to (1/2) \hat{I}$, while the distribution of states $P_\infty(\theta)$ is highly non-trivial. The mean number of detector clicks and the complete distribution $P(\theta,t)$ is experimentally accessible using the methods of quantum tomography and our results could be experimentally verified. Finally, we hope that the connection between  the stochastic Schr\"odinger equation and resetting dynamics, pointed out in this work, will lead to further useful results and insights in both areas.

\ack
We thank Cedric Bernardin for valuable discussions. AD thanks  Sarang Gopalakrishnan, David Huse and Anupam Kundu for useful discussions. VD thanks Cl\'ement Pellegrini and Vishal Vasan for valuable discussions on counting processes and spectral analysis. VD  thanks Gregory Schehr and  Satya Majumdar for useful discussions on the initial draft of this work. VD and AD acknowledge the support of the DAE, Govt. of India, under project no. 12-R\&D-TFR-5.10-1100 and project no. RTI4001. RC is supported by the French National Research Agency through the projects QTraj (ANR-20-CE40-0024-01), RETENU (ANR-20-CE40-0005-01), and ESQuisses (ANR-20-CE47- 0014-01). The numerical calculations were performed on the computational facilities at ICTS.

\subsection*{Data Availability Policy}
The authors confirm that any data that support the findings of this study are included within the article.

\appendix

\section{Computation of survival probability}
\label{sec:CalSurvProb}
We use the notation introduced below Eq.~(\ref{eq:SurvProb}). When $0 \le \lambda < 1$, then $H_{\mathit{eff}}$ admits the orthogonal decomposition
\begin{equation*}
\frac{1}{2\beta}
\left[\begin{array}{cc}
c & c^{*} \\
-c^{*} & c
\end{array}\right]
\left[\begin{array}{cc}
-(c^{*})^2 & 0 \\
0 & c^2 
\end{array}\right]
\left[\begin{array}{cc}
c & -c^{*} \\
c^{*} & c
\end{array} \right]
\end{equation*}
where $c=\exp\left[\rmi(\frac{2\phi-\pi}{4})\right]$. From the above and the fact that $c^{2}+(c^{*})^{2}=2\beta$, one has
\begin{equation*}
\fl \rme^{-\rmi \gamma_{0} t H_{\mathit{eff}}}= \frac{1}{2\beta}\left[\begin{array}{cc}
c & c^{*} \\
-c^{*} & c
\end{array}\right]
\left[\begin{array}{cc}
\exp[\rmi\gamma_{0}t(c^{*})^2] & 0 \\
0 & \exp[-\rmi\gamma_{0}t c^2] 
\end{array}\right]
\left[\begin{array}{cc}
c & -c^{*} \\
c^{*} & c
\end{array} \right].
\end{equation*}
With the initial condition $\psi(0)=\psi_{0}$, from Eq.~(\ref{eq:NoClickNonHerm}) one has
\begin{equation}
\tilde{\psi}(t)=\rme^{-\rmi \gamma_{0} t H_{\mathit{eff}}}\psi_{0} \nonumber = \frac{1}{\beta}
\left[\begin{array}{c}
{\rm Re} [c^2 \exp(\rmi \gamma_{0}t (c^{*})^2)]\\
-\rmi\, {\rm Im} [\exp( \rmi \gamma_{0}t(c^{*})^2)]
\end{array}\right].
\label{eq:danser}
\end{equation}
The formula for survival probability now follows from the definition in Eq.~(\ref{eq:DefSu}). The calculation for $\lambda>1$ is similar.

For $\lambda=1$, as noted earlier $H_{\mathit{eff}}$ is non-diagonalizable. The Jordan decomposition of $H_{\mathit{eff}}$ is
\begin{equation*}
\left[\begin{array}{cc}
1 & \rmi \\
0 & 1
\end{array}\right]
\left[\begin{array}{cc}
-\rmi & 0 \\
1 & -\rmi
\end{array}\right]
\left[\begin{array}{cc}
1 & -\rmi \\
0 & 1
\end{array}\right].
\end{equation*}
Once again, matrix exponentiation gives
\begin{equation*}
\rme^{-\rmi \gamma_{0} t H_{\mathit{eff}}}= \rme^{-\gamma_{0}t}
\left[\begin{array}{cc}
1+\gamma_{0}t & -\rmi\gamma_{0}t \\
-\rmi\gamma_{0}t &  1-\gamma_{0}t
\end{array}\right].
\end{equation*}
With the same initial condition as before, one obtains
\begin{equation}
\tilde{\psi}(t)= \rme^{-\gamma t/4}
\left[\begin{array}{c}
1+\gamma_{0} t \\
-\rmi \gamma_{0} t\\
\end{array}\right]
\label{eq:danser2}
\end{equation}
and the expression for $S(t)$ follows.

\section{Evolution of a qubit on the Bloch sphere}
\label{sec:QubitEvol}
The standard representation of a qubit state on the Bloch sphere is given by
\begin{equation*}
|\psi\ra=\cos\frac{\chi}{2} |\psi_{0}\ra+\rme^{\rmi \xi}\sin\frac{\chi}{2}|\psi_{1}\ra
\end{equation*}
for $0\le\chi\le\pi$ and $0\le\xi\le 2\pi$. In the spherical polar coordinate system, $\chi$ is the polar angle and $\xi$ the azimuthal angle. In the $yz$ plane, $\xi=\pi/2$ for $y>0$ and $\xi=3\pi/2$ for $y<0$. $\xi$ is undefined on the $z$ axis. Substituting the above in Eq.~(\ref{eq:NoClickNonLin}), the following coupled system is obtained
\begin{eqnarray*}
\sin\frac{\chi}{2}\,\left[\rmi\dot{\chi}+2\gamma_{0}(\rmi\lambda\sin\chi + \rme^{\rmi\xi})\right]=0, \\
\cos\frac{\chi}{2}\,\left[\rmi\dot{\chi}-2\gamma_{0}(-\rmi\lambda\sin\chi + \rme^{-\rmi\xi})\right]-2\dot{\xi}\,\sin\frac{\chi}{2}=0.
\end{eqnarray*}
The trajectory of constant $\xi$ is for $\cos\xi_{0}=0$ which corresponds to $\xi_{0}=\pi/2$ and $\xi_{0}=3\pi/2$.  For $\xi_{0}=\pi/2$, $\chi$ evolves in accordance with
\begin{equation*}
\dot{\chi}=-2\gamma_{0}(\lambda\sin\chi + 1),
\end{equation*}
whereas for $\xi_{0}=3\pi/2$, $\chi$ evolves in accordance with
\begin{equation*}
\dot{\chi}=-2\gamma_{0}(\lambda\sin\chi - 1).
\end{equation*}
On the part of the evolution for $\xi=\pi/2$, define $\theta=\chi$ and on the part of the evolution for $\xi=3\pi/2$, define $\theta=-\chi$. Then the entire evolution of the state vector in Eq.~(\ref{eq:GenStateYZ}) happens via Eq.~(\ref{eq:AngularSpeed}).

\section{ Solution for the flow of Eq.~(\ref{eq:AngularSpeed})}
\label{sec:integTheta}
Here we note  the results of integration of  Eq.~(\ref{eq:AngularSpeed}) for the various cases of $\lambda$.

\begin{itemize}
\item For $\lambda<1$, $\theta_{0}=0$, the equation integrates to give
\begin{equation}
\label{eq:alphaLT2}
\arctan\frac{\lambda+\tan[\theta_{t}(0,0)/2]}{\sqrt{1-\lambda^{2}}} - \arctan\frac{\lambda}{\sqrt{1-\lambda^{2}}} = -\beta\gamma_{0}t.
\end{equation}
If a jump occurs at $t_{1}$, then $\theta_{t_{1}}=\pi$ and  evolution happens via no click from $t_{1}$ to $t$ then
\begin{equation}
\label{eq:alphaLTz}
\arctan\frac{\lambda+\tan[\theta_{t}(t_{1},\pi)/2]}{\sqrt{1-\lambda^{2}}} - \frac{\pi}{2} = -\beta\gamma_{0}(t-t_{1}).
\end{equation}

\item For $\lambda=1$ and $\theta(0)=0$, the equation  integrates to give
\begin{equation}
\label{eq:alphaEQ2}
\tan\left(\frac{\pi}{4}-\frac{\theta_{t}(0,0)}{2}\right)=1+2\gamma_{0} t.
\end{equation}
For $\lambda=1$, if a jump occurs at $t_{1}$, then $\theta_{t_{1}}=\pi$ and  evolution happens via no click from $t_{1}$ to $t$. The equation integrates to give
\begin{equation}
\label{eq:alphaEQ2z}
\tan\left(\frac{\pi}{4}-\frac{\theta_{t}(t_{1},\pi)}{2}\right)=-1+2\gamma_{0} (t-t_{1}).
\end{equation}

\item For $\lambda>1$ and $\theta(0)=0$, the equation  integrates to give
\begin{equation}
\label{eq:alphaGT2}
\tan\frac{\theta_{t}(0,0)}{2}=-\frac{\sinh(\beta'\gamma_{0}t)}{\sinh(\beta'\gamma_{0}t+\phi')}.
\end{equation}
If a jump occurs at $t_{1}$, then $\theta_{t_{1}}=\pi$ and  evolution happens via no click from $t_{1}$ to $t$ then
\begin{equation}
\label{eq:alphaGT2z}
\tan\frac{\theta_{t}(t_{1},\pi)}{2}=-\frac{\sinh(\beta'\gamma_{0}(t-t_{1})-\phi')}{\sinh(\beta'\gamma_{0}(t-t_{1}))}.
\end{equation}
\end{itemize}

\section{Laplace Transform of the moment generating function}
\label{sec:MGFCalc}

Consider the case $\lambda<1$. Define the function
\begin{equation*}
g(t,\phi)=\sin^{2}(\beta\gamma_{0}t-\phi)
\end{equation*} 
so that from Eqs.~(\ref{eq:NClickDensity} , \ref{eq:alphaLT2}, \ref{eq:alphaLTz} and \ref{eq:nCountProb}), one has
\begin{equation}
\label{eq:nCountConv}
\eqalign{P_{0}^{t}[0]=\frac{\rme^{-\frac{\gamma t}{2}}}{\beta^{2}}\left[g(t,0)+g(t,-\phi)\right], \\
\fl	P_{0}^{t}[n]=\frac{\gamma^{n}\rme^{-\frac{\gamma t}{2}}}{\beta^{2n+2}}\int_{0}^{t}dt_{n}\left[g(t-t_{n},0)+g(t-t_{n},\phi)\right]\left[\prod_{k=n-1}^{1}\int_{0}^{t_{k+1}}dt_{k}\,g(t_{k+1}-t_{k},\phi)\right]g(t_{1},0).}
\end{equation} 
Notice $P_{0}^{t}[n]$ has been expressed as a convolution. 
For the function $g(t,\phi)$, the Laplace transform is
\begin{equation}
\label{eq:LTransOfg}
\eqalign{ \hat{g}_{\phi}(\sigma)&=\int_{0}^{\infty} \rme^{-\left(\sigma+\frac{\gamma}{2}\right)t} g(t,\phi)dt \\&=
\frac{1}{2}\left[\frac{1}{\sigma+\frac{\gamma}{2}}-\frac{(\sigma+\frac{\gamma}{2})\cos(2\phi)+2\beta\gamma_{0}\sin(2\phi)}{(\sigma+\frac{\gamma}{2})^{2}+4\beta^2\gamma_{0}^2}\right].}
\end{equation}
From Eqs.~(\ref{eq:MGF}, \ref{eq:nCountConv}) and standard properties of Laplace transform, one has
\begin{equation*}
\eqalign{(\mathfrak{L}\mathbb{E}[\rme^{-s N_{t}}])(\sigma,s)&=\int_{0}^{\infty}\rme^{-\sigma t}\left(\sum_{n\ge 0}\rme^{-n s}P_{0}^{t}[n]\right)\,dt\\
	&=\frac{1}{\beta^2}\left[\hat{g}_{-\phi}+\hat{g}_{0}+\frac{\gamma \rme^{-s}}{\beta^{2}}\frac{\hat{g}_{0}[\hat{g}_{0}+\hat{g}_{\phi}]}{1-\frac{\gamma \rme^{-s}}{\beta^{2}}\hat{g}_{\phi}}\right].}
	\end{equation*}
%
From Eq.~(\ref{eq:LTransOfg}) and the above, Eq.~(\ref{eq:LTransMGF}) follows after simplification. The calculations for $\lambda=1$ and $\lambda>1$ are similar and lead to the same Eq.~(\ref{eq:LTransMGF}).


\vskip 0.5cm

{\bf References} \vskip 0.5cm
\bibliographystyle{iopart-num}
\bibliography{references}

\end{document}